\documentclass{jaa}
\usepackage{natbib}
\usepackage{graphicx}
\usepackage{amsmath}    
\usepackage{amssymb}   
\usepackage{subcaption}
\usepackage{graphicx}

\begin{document}\sloppy

\title{On the origin of transient features in cosmological N-Body Simulations}

\author{J.~S.~Bagla\textsuperscript{1,2,*} and Swati Gavas\textsuperscript{1}}
\affilOne{\textsuperscript{1}Department of Physical Sciences, Indian Institute of Science Education and Research Mohali, Sector 81, S.A.S. Nagar, Punjab 140306, India. \\}
\affilTwo{\textsuperscript{2}National Centre for Radio Astrophysics, Tata Institute of Fundamental Research, Ganeshkhind, Pune 411007, India}

\twocolumn[{

\maketitle

\corres{jasjeet@iisermohali.ac.in}

\msinfo{16 August 2024}{29 January 2025}

%%abstract
\begin{abstract}
We study the effect of gravitational clustering at small scales on larger scales by studying mode coupling between virialised halos. We build on the calculation by Peebles (1974) where it was shown that a virialised halo does not contribute any mode coupling terms at small wave numbers $k$. Using a perturbative expansion in wave number, we show that this effect is small and arises from the deviation of halo shapes from spherical and also on tidal interactions between halos. We connect this with the impact of finite mass resolution of cosmological N-Body simulations on the evolution of perturbations at early times. This difference between the expected evolution and the evolution obtained in cosmological N-Body simulations can be quantified using such an estimate. We also explore the impact of a finite shortest scale up to which the desired power spectrum is realised in simulations. Several simulation studies have shown that this effect is small in comparison with the effect of perturbations at large scales on smaller scales. It is nevertheless important to study these effects and develop a general approach for estimating their magnitude. This is especially relevant in the present era of precision cosmology. We provide basic estimates of the magnitude of these effects and their power spectrum dependence. We find that the impact of small scale cutoff in the initial power spectrum and discreteness increases with $(n+3)$, with $n$ being the index of the power spectrum. In general, we recommend that cosmological simulation data should be used only if the scale of non-linearity, defined as the scale where the linearly extrapolated {\it rms} amplitude of fluctuations is unity, is larger than the average inter-particle separation.
\end{abstract}


}]

\doinum{12.3456/s78910-011-012-3}
\artcitid{\#\#\#\#}
\volnum{000}
\year{0000}
\pgrange{1--}
\setcounter{page}{1}
\lp{1}

\section{Introduction}
It is believed that structures like galaxies and clusters of galaxies formed by the growth of small perturbations due to gravitational clustering \citep{peebles1981, shandarin1989, peacock1998, bernardeau2001, padmanabhan2002}. 
The same processes are thought to be responsible for the formation of the large scale structure, i.e., the clustered distribution of galaxies in space. 
Dark matter is believed to be the dominant form of matter in galaxies and clusters of galaxies: dark matter is generally assumed to be made up of non-relativistic, weakly interacting massive particles (WIMPs) \citep{trimble1987, spergel2003,planck2020}. 
In view of the weak interactions within the dark matter sector, it responds mainly to gravitational forces.
As dark matter dominates over baryonic matter in terms of total matter content, assembly of matter into halos and the large scale structure is driven mainly by gravitational amplification of initial perturbations. 
In the standard cosmological model, it is believed that galaxies form when gas in highly over-dense halos forms stars due to cooling, collapse and fragmentation of gas \citep{hoyle1953, rees1977, silk1977, binney1977, benson2010, somerville2015,wechsler2018}. 
Thus, the evolution of density perturbations due to gravitational clustering in an expanding universe is a key process for the study of large scale structure and its evolution. 
The basic equations for this can be derived from first principles \citep{peebles1974}.
These equations can be solved easily when the amplitude of perturbations is small. 
At this stage, perturbations at each scale evolve independently, and mode coupling is sub-dominant. 
Once the amplitude of perturbations at relevant scales becomes large, the coupling with perturbations at other scales becomes important and cannot be ignored, though a few results have been obtained. 
The equation for the evolution of density perturbations cannot be solved for generic perturbations in the non-linear regime. 
One can use dynamical approximations for studying mildly non-linear perturbations
\citep{zeldovich1970, gurbatov1989, matarrese1992, brainerd1993, bagla1994, sahni1995, hui1996, bernardeau2001,crocce2006,kitaura2013,pixius2022, ota2022, garny2023, garny2023a, rampf2023}. 
Scaling relations can be used if the aim is to describe the system using statistical descriptors  \citep{davis1977, hamilton1991, jain1995, kanekar2000, ma1998, nityananda1994, padmanabhan1995, padmanabhan1996, peacock1996, smith2003,carrasco2012,baumann2012}.
We require cosmological N-Body simulations \citep{bagla1994, bertschinger1998, bagla2004,angulo2022} to follow the detailed evolution of generic systems.

Cosmological N-Body simulations work with a representative region of the universe. 
This region is typically large but finite. 
Perturbations at scales smaller than the mass resolution of the simulation, and at scales larger than the box cannot be represented or taken into account.  
Indeed, perturbations at scales comparable to the box are under-sampled with only a handful of independent modes.  
Perturbations at scales much larger than the simulation volume can affect the results of N-Body simulations  \citep{gelb1994, gelb1994a, tormen1996, cole1997, bagla2005, bagla2006, power2006, takahashi2008, bagla2009, angulo2016, michaux2020}. 
It is possible to estimate whether a given simulation volume is large enough to be representative or not
\citep{bagla2005, bagla2006}. 
It has also been shown that for gravitational dynamics in an expanding universe, perturbations at small scales do not influence the collapse of large scale perturbations in a significant manner if the scale of non-linearity is larger than the smaller scales under consideration  \citep{peebles1974, peebles1985, little1991, bagla1997b, couchman1998}.
This is certainly valid for the correlation function or power spectrum at large scales. 
This has led to a belief that ignoring perturbations at scales much smaller than the scales of interest does not affect the results of N-Body simulations. 
In earlier work, we have shown that if large scale collapse is highly symmetric, then the presence of perturbations at much smaller scales affects the evolution of density perturbations at large scales \citep{bagla2005a}. 
However, such effects are not obvious when we study the evolution of perturbations with generic initial conditions.

Substructure is known to play an important role in the dynamical relaxation of halos. 
It can induce mixing in phase space \citep{lynden-bell1967, weinberg2001}, or change halo profiles by introducing transverse motions \citep{peebles1990, subramanian2000}.
Further, gravitational interactions between substructures can introduce an effective collisionality even for a collisionless system \citep{ma2004, ma2003}. 
It is, therefore, important to understand the role played by substructure in gravitational clustering and dynamical relaxation in the cosmological context.

The key mechanism for perturbations across scales affecting each other is mode coupling.  
We know the following from earlier work. 
\begin{itemize}
\item
Perturbations at large scales influence perturbations at small scales in a significant manner. 
If the initial conditions are modified by filtering out perturbations at small scales or other modifications restricted to small scales, then mode coupling generates small scale power. 
If the scale of filtration is smaller than the scale of non-linearity at the final epoch, then the non-linear power spectrum as well as the appearance of large scale structure is similar to the original initial conditions without modification 
\citep{peebles1985, little1991, bagla1997b, couchman1998}.
\item
Non-linear evolution of density perturbations {\it drives} every model towards a weak attractor ($P(k)\simeq k^{-1}$) in the mildly non-linear regime ($1 \leq \bar\xi \leq 200$) \citep{klypin1992,bagla1997b}. 
\item
If there are no initial perturbations at large scales, mode coupling generates power with ($P(k) \simeq k^4$) that grows very rapidly at early times \citep{bagla1997b}.  
This can be explained in many ways.  
The second-order perturbation theory as well as momentum and mass conserving motion of a group of particles provide an adequate explanation.  
The $k^4$ tail can also be derived from the full non-linear equation for density \citep{peebles1974, peebles1981, zeldovich1965}.
\item
In case the large scale perturbations are highly symmetric, e.g. planar, then small scale fluctuations play an important role in the non-linear evolution of perturbations at large scales \citep{bagla2005a}. 
\end{itemize}

The effect of large scales on small scales is clearly significant, particularly if the larger scales are comparable to the scale of non-linearity.
On the other hand, the effect of small scales on larger scales is known to be small in most situations. 
Though this effect has not been studied in detail, many tools have been developed that exploit the presumed smallness of the influence of small scales
on large scales \citep{bond1996, monaco2002, monaco2002a}. 

{\sl Pre-initial}\/ conditions refer to the distribution of particles that is laid out before imposing initial density and velocity perturbations. 
The choice of pre-initial conditions is known to impact the results \citep{baertschiger2007, baertschiger2007a, baertschiger2007b, bagla1997, gabrielli2006, joyce2007, joyce2007a, joyce2005, marcos2006,lhuillier2014,michaux2020,zhang2021}. 
Pre-initial conditions are expected to have no density perturbations or symmetry. 
It is clear, however, that at least one of these requirements must be relaxed in practice. 
This leads to growth of some modes deviating from expectations in perturbation theory. 
The work presented here allows us to estimate the effect such discrepant modes can have on the non-linear evolution of clustering across scales. 

The representation and evolution of perturbations at small scales depend strongly on the mass and force resolution in the simulation. 
A high force resolution without a matching mass resolution can lead to better modelling of dense halos but gives rise to two body collisions and misleading results in some situations  \citep{kuhlman1996,splinter1998, binney2002, diemand2004, binney2004, el-zant2006, romeo2008,mansfield2020}. 
Further, discreteness and stochasticity limit our ability to measure physical quantities in simulations \citep{thiebaut2008, romeo2008,angulo2022}. 
The question we address here is: can these errors remain significant at late times, and, can these errors propagate to larger scales through mode coupling?
Propagation of these errors can potentially introduce transient features in N-Body simulations that are likely to remain dominant till the mass scale of non-linearity is well resolved. 

We present our analysis in two parts. In \S{2}, we discuss mode coupling and the expected effect of collapsed halos at small scales on perturbations at much larger scales. In \S{3}, we present an analysis of the effect of unrepresented perturbations at small scales. We also present an analysis of the same using N-Body simulations. This is followed by discussion and summary in \S{4}.

\section{Mode Coupling}

The evolution of density perturbations in a system of particles interacting gravitationally in an expanding universe \citep{peebles1981} can be written as:
\begin{align}
{\ddot\delta}_{\mathbf k} + 2 \frac{\dot{a}}{a} {\dot\delta}_{\mathbf k} &=A_{\mathbf k} - B_{\mathbf k} \\
A_{\mathbf k} &= \frac{1}{M}\sum\limits_{j} m_j \left [ \iota {\mathbf k}. \left( - \frac{\mathbf \nabla \phi_j}{a^2} \right) \exp\left(\iota {\mathbf  k}.{\mathbf x}_j \right) \right ] \\
B_{\mathbf k} &= \frac{1}{M} \sum\limits_{j} m_j \left({\mathbf k}.{\dot {\mathbf x}_j } \right)^2   \exp\left[\iota {\mathbf  k}.{\mathbf x}_j \right] 
\end{align}
The usual linear term in $\delta_{\mathbf k}$ is absorbed in $A_{\mathbf k}$. It has been shown that a single cluster of particles in virial equilibrium does not contribute to $A_{\mathbf k} - B_{\mathbf k}$ at wave numbers much smaller than the inverse of the cluster size. Here, we consider the effect of interaction between two clusters in order to estimate the effect of mode coupling and transfer of power from small scales to large scales.

The first mode coupling term can be written as:
\begin{align}
A_{\mathbf k} &= \frac{1}{M}\sum\limits_{j} m_j \left [ \iota  {\mathbf k}. \left( - \frac{\mathbf \nabla \phi_j}{a^2} \right) \exp\left(\iota {\mathbf  k}.{\mathbf x}_j \right) \right ] \nonumber \\ 
&= \frac{\iota }{M} \sum\limits_{j} \left({\mathbf k}.{\mathbf f}_j \right) \exp\left(\iota {\mathbf  k}.{\mathbf x}_j \right)  
\end{align}
if there are two clusters $C_1$ and $C_2$ then the sum in RHS can be written separately for particles in two clusters.
\begin{equation}
A_{\mathbf k} = \frac{\iota }{M} \sum\limits_{j\in C_1} 
 \left({\mathbf k}.{\mathbf f}_j \right) \exp\left(\iota {\mathbf  k}.{\mathbf x}_j \right)  + \frac{\iota }{M} \sum\limits_{j \in C_2}  \left({\mathbf k}.{\mathbf f}_j \right) \exp\left(\iota {\mathbf  k}.{\mathbf x}_j \right) 
\end{equation}
representing the force acting on particles in halo '1' due to halo '2' by ${\mathbf f}^{12}$, and the force due to particles within the same halo by ${\mathbf f}^{11}$, we can rewrite the sum\footnote{Note that there is no self-interaction implied, each particle experiences a force due to all the others.} as: 
\begin{align}
A_{\mathbf k} &= \frac{\iota }{M} \sum\limits_{j \in C_1} \left({\mathbf k}.{\mathbf f}_j^{11} \right) \exp\left(\iota {\mathbf k}.{\mathbf x}_j \right) \nonumber \\
& \quad + \frac{\iota }{M} \sum\limits_{j \in C_1} \left({\mathbf k}.{\mathbf f}_j^{12} \right) \exp\left(\iota {\mathbf k}.{\mathbf x}_j \right)  \nonumber \\
& \quad + \frac{\iota }{M} \sum\limits_{j \in C_2} \left({\mathbf k}.{\mathbf f}_j^{22} \right) \exp\left(\iota {\mathbf k}.{\mathbf x}_j \right) \nonumber\\
& \quad +  \frac{\iota }{M} \sum\limits_{j \in C_2} \left({\mathbf k}.{\mathbf f}_j^{21} \right) \exp\left(\iota {\mathbf k}.{\mathbf x}_j \right)  
\end{align}
If the wave numbers of interest are such that $\left|{\mathbf k}. {\mathbf r} \right|  \ll 1$, then we can expand the exponential in each term.
\begin{align}
A_{\mathbf k} &= \frac{\iota }{M} \sum\limits_{j \in C_1}  \left({\mathbf k}.{\mathbf f}_j^{11} \right) \left[ 1 + \iota  \left({\mathbf k}. {\mathbf x}_j \right) + \mathcal{O}(k^2) \right] \nonumber\\
& \quad + \frac{\iota }{M} \sum\limits_{j \in C_1}  \left({\mathbf k}.{\mathbf f}_j^{12} \right) \left[ 1 + \iota  \left({\mathbf k}.{\mathbf x}_j \right) + \mathcal{O}(k^2) \right] \nonumber \\
& \quad + \frac{\iota }{M} \sum\limits_{j \in C_2}  \left({\mathbf k}.{\mathbf f}_j^{22} \right) \left[ 1 + \iota  \left({\mathbf k}.{\mathbf x}_j \right) +   \mathcal{O}(k^2) \right] \nonumber\\
& \quad + \frac{\iota }{M} \sum\limits_{j \in C_2}  \left({\mathbf k}.{\mathbf f}_j^{21} \right) \left[ 1 + \iota  \left({\mathbf k}.{\mathbf x}_j \right) +  \mathcal{O}(k^2) \right] \nonumber \\
&= \frac{\iota }{M} {\mathbf k}. \left[ \sum\limits_{j \in C_1} {\mathbf f}_j^{11} + \sum\limits_{j \in C_1} {\mathbf f}_j^{12} + \sum\limits_{j \in C_2} {\mathbf f}_j^{22} + \sum\limits_{j \in C_2} {\mathbf f}_j^{21} \right] \nonumber \\
& \quad -  \frac{1}{M} {\mathbf k}. \left[ \sum\limits_{j \in C_1} {\mathbf f}_j^{11} \left( {\mathbf k}.{\mathbf x}_j \right) + \sum\limits_{j \in C_1} {\mathbf f}_j^{12} \left( {\mathbf k}.{\mathbf x}_j \right) \right. \nonumber\\
& \qquad \qquad \left.+ \sum\limits_{j \in C_2} {\mathbf f}_j^{22} \left( {\mathbf k}.{\mathbf x}_j \right) + \sum\limits_{j \in C_2} {\mathbf f}_j^{21} \left( {\mathbf k}{\mathbf x}_j
\right)  \right] \nonumber \\
& \quad + \mathcal{O}(k^3) 
\end{align}
The first and the third term at $\mathcal{O}(k)$ are zero as there is no net internal force. Second and fourth terms at the same order cancel by virtue of Newton's third law. Thus, there is no contribution at this order and the lowest order contribution is at $\mathcal{O}(k^2)$ from this term. The external force may be assumed to be constant across the cluster\footnote{In principle, we can evaluate the tidal interaction term as well, but that is subdominant in most situations, and we choose to ignore it here.}. This allows us to simplify the second and fourth terms.
\begin{align}
A_{\mathbf k} &\simeq - \frac{1}{M} {\mathbf k}. \left[ \sum\limits_{j \in C_1} {\mathbf f}_j^{11} \left( {\mathbf k}.{\mathbf x}_j \right) + \sum\limits_{j \in C_1} {\mathbf f}_j^{12} \left( {\mathbf k}.{\mathbf x}_j \right) \right. \nonumber \\
& \qquad \qquad \left.+ \sum\limits_{j \in C_2} {\mathbf f}_j^{22} \left( {\mathbf k}.{\mathbf x}_j \right) + \sum\limits_{j \in C_2} {\mathbf f}_j^{21} \left( {\mathbf k}.{\mathbf x}_j
\right)  \right] \nonumber \\
&\simeq - \frac{1}{M} {\mathbf k}. \left[ \sum\limits_{j \in C_1} {\mathbf f}_j^{11} \left( {\mathbf k}.{\mathbf x}_j \right)
+  {\mathbf F}^{12} \left( {\mathbf k}.{\mathbf X}_1 \right) \right. \nonumber \\ 
& \qquad \qquad \left.+ \sum\limits_{j \in C_2} {\mathbf f}_j^{22} \left( {\mathbf k}.{\mathbf x}_j \right) + {\mathbf F}^{21} \left( {\mathbf k}.{\mathbf X}_2 \right)  \right]   \nonumber \\
&= - \frac{1}{M} {\mathbf k}. \left[ \sum\limits_{j \in C_1} {\mathbf
f}_j^{11} \left( {\mathbf k}.{\mathbf x}_j \right) + \sum\limits_{j \in C_2} {\mathbf f}_j^{22} \left( {\mathbf k}.{\mathbf x}_j
\right) \right] \nonumber \\
& \quad-  \frac{1}{M} {\mathbf k}.{\mathbf F}^{12} \left( {\mathbf k}.\left({\mathbf X}_1 - {\mathbf  X}_2 \right) \right) 
\end{align}
Here ${\mathbf F}^{ij}$ is the force due to the $j$th cluster in the $i$th cluster, and ${\mathbf X}_i$ is the centre of mass of the $i$th cluster. It is easy to see that the combination of reduced terms scales inversely with the separation of the two clusters. Terms that refer to the internal forces can also be rewritten by switching to the centre of mass frame of each cluster. We use centre of mass coordinates ${\mathbf y}_j^i = {\mathbf x}_j - {\mathbf
X}_i$ for the $j$th particle in the $i$th cluster. We will drop the superscript as it is obvious from the context.
\begin{align}
A_{\mathbf k} &= - \frac{1}{M} {\mathbf k}. \left[ \sum\limits_{j \in C_1} {\mathbf f}_j^{11} \left( {\mathbf k}.\left({\mathbf y}_j + {\mathbf X}_1\right)  \right) \right. \nonumber \\
& \qquad \qquad \left. + \sum\limits_{j \in C_2} {\mathbf f}_j^{22} \left( {\mathbf k}.\left({\mathbf y}_j + {\mathbf X}_2\right) \right)  \right] \nonumber \\
& \quad-  \frac{1}{M} {\mathbf k}.{\mathbf F}^{12} \left( {\mathbf k}.\left({\mathbf X}_1 - {\mathbf  X}_2 \right) \right) 
\nonumber \\
&= - \frac{1}{M} {\mathbf k}. \left[ \sum\limits_{j \in C_1} {\mathbf
f}_j^{11} \left( {\mathbf k}.{\mathbf y}_j  \right) + \sum\limits_{j \in C_2} {\mathbf f}_j^{22} \left( {\mathbf k}.{\mathbf y}_j \right)  \right] \nonumber \\
&\quad -  \frac{1}{M} {\mathbf k}.{\mathbf F}^{12} \left( {\mathbf k}.\left({\mathbf X}_1 - {\mathbf  X}_2 \right) \right) + \mathcal{O}(k^3)
\end{align}
The second step follows as the internal forces cancel out for each cluster. This is the final expression for $A_{\mathbf k}$, ignoring the tidal interaction terms between the two clusters.

Now, we consider the second mode coupling term.
\begin{equation}
B_{\mathbf k} = \frac{1}{M} \sum\limits_{j} m_j \left({\mathbf k}.{\dot {\mathbf x}_j } \right)^2   \exp\left[\iota {\mathbf  k}.{\mathbf x}_j \right]
\end{equation}
It is evident that this term will contribute at $\mathcal{O}(k^2)$ or higher orders to the evolution of $\delta_{\mathbf k}$. We can again write the sum in two parts, one for each cluster.
\begin{align}
B_{\mathbf k} &= \frac{1}{M} \sum\limits_{j} m_j \left({\mathbf k}.{\dot {\mathbf x}_j } \right)^2   \exp\left[\iota {\mathbf  k}.{\mathbf x}_j \right] \nonumber \\
&= \frac{1}{M} \sum\limits_{j \in C_1}  m_j \left({\mathbf k}.{\dot
{\mathbf x}_j } \right)^2   \exp\left[\iota {\mathbf  k}.{\mathbf x}_j \right] \nonumber \\
& \quad + \frac{1}{M}  \sum\limits_{j \in C_2}  m_j \left({\mathbf k}.{\dot {\mathbf x}_j } \right)^2   \exp\left[\iota {\mathbf  k}.{\mathbf x}_j \right] \nonumber \\
&\simeq
\frac{1}{M}  \sum\limits_{j \in C_1}  m_j \left({\mathbf k}.{\dot
    {\mathbf x}_j } \right)^2
+ \frac{1}{M} \sum\limits_{j \in C_2}  m_j \left({\mathbf k}.{\dot
    {\mathbf x}_j } \right)^2  + \mathcal{O}(k^3)
\end{align}
We can now switch to the centre of mass frame for each cluster. Let ${\mathbf V}$ be the velocity of the centre of mass and $\dot{{\mathbf y}}_j$ the velocity of the $j$th particle in this frame.
\begin{align}
B_{\mathbf k} &= \frac{1}{M}  \sum\limits_{j \in C_1}  m_j \left[{\mathbf k}.\left({\dot {\mathbf y}_j} + {\mathbf V}_1\right) \right]^2 \nonumber \\
& \quad + \frac{1}{M} \sum\limits_{j \in C_2}  m_j \left[{\mathbf k}.\left({\dot {\mathbf y}_j} + {\mathbf V}_2\right)\right]^2  + \mathcal{O}(k^3) \nonumber \\
&= \frac{1}{M}  \sum\limits_{j \in C_1}  m_j \left[\left({\mathbf
k}.{\dot {\mathbf y}_j}\right)^2 + \left({\mathbf k}.{\mathbf
V}_1\right)^2 + 2 \left({\mathbf k}.{\dot {\mathbf y}_j}\right)
\left({\mathbf k}.{\mathbf  V}_1\right) \right]\nonumber \\
&\quad +  \frac{1}{M}  \sum\limits_{j \in C_2}  m_j \left[\left({\mathbf k}.{\dot {\mathbf y}_j}\right)^2 + \left({\mathbf k}.{\mathbf V}_2\right)^2 + 2 \left({\mathbf k}.{\dot {\mathbf y}_j}\right) \left({\mathbf k}.{\mathbf  V}_2\right) \right] \nonumber \\
& \quad + \mathcal{O}(k^3) \nonumber \\
&= \frac{1}{M}  \sum\limits_{j \in C_1}   m_j \left({\mathbf
k}.{\dot {\mathbf y}_j}\right)^2 +  \frac{1}{M}  \sum\limits_{j \in C_2}   m_j \left({\mathbf k}.{\dot {\mathbf y}_j}\right)^2 \nonumber \\
&  \quad + \frac{1}{M}  \left[ M_1 \left({\mathbf k}.{\mathbf V}_1\right)^2 + M_2 \left({\mathbf k}.{\mathbf V}_2\right)^2 \right]  + \mathcal{O}(k^3)  
\end{align}
The total contribution to mode coupling can be written as:
\begin{align}
A_{\mathbf k} - B_{\mathbf k} &= 
- \frac{1}{M} {\mathbf k}. \left[ \sum\limits_{j \in C_1} {\mathbf
f}_j^{11} \left( {\mathbf k}.{\mathbf y}_j  \right) + \sum\limits_{j \in C_2} {\mathbf f}_j^{22} \left( {\mathbf k}.{\mathbf y}_j \right)  \right] \nonumber \\
& \quad +  \frac{1}{M}  \sum\limits_{j \in C_1}   m_j \left({\mathbf
k}.{\dot {\mathbf y}_j}\right)^2 +  \frac{1}{M}  \sum\limits_{j \in C_2}   m_j \left({\mathbf k}.{\dot {\mathbf y}_j}\right)^2 
\nonumber \\
& \quad -  \frac{1}{M} {\mathbf k}.{\mathbf F}^{12} \left( {\mathbf k}.\left({\mathbf X}_1 - {\mathbf  X}_2 \right) \right) \nonumber \\
& \quad+ \frac{1}{M}  \left[ M_1 \left({\mathbf k}.{\mathbf V}_1\right)^2 + M_2 \left({\mathbf k}.{\mathbf V}_2\right)^2 \right]  
\nonumber \\
& \quad + \mathcal{O}(k^3) 
\end{align}
Using the fact that $A_{\mathbf k} - B_{\mathbf k}$ for a virialised cluster, computed in its centre of mass is zero at this order, we find:
\begin{align}
A_{\mathbf k} - B_{\mathbf k} &= - \frac{1}{M} {\mathbf k}.{\mathbf F}^{12} \left( {\mathbf k}.\left({\mathbf X}_1 - {\mathbf  X}_2 \right) \right) \nonumber \\
& \quad + \frac{1}{M}  \left[ M_1 \left({\mathbf k}.{\mathbf V}_1\right)^2 + M_2 \left({\mathbf k}.{\mathbf V}_2\right)^2 \right]  + \mathcal{O}(k^3) 
\end{align}
Thus, the contribution of interacting clusters to mode coupling leads to the influence of small scales on larger scales, which can be thought of in terms of clusters acting as point masses to the leading order. Here, we have ignored the subdominant tidal interactions between the clusters. Both the leading and subleading terms lead to the generation of a power spectrum that goes as $k^4$ at small $k$. This expression encodes the fact that the internal structure of virialised interacting clusters at small scales does not affect the evolution of perturbations at much larger scales to the leading order. This equation can be generalised to an arbitrary number of interacting clusters. 

Going further with this analysis, especially to put these in the context of specific models is non-trivial.  
However, the message to take home from here is that there is a weak coupling between non-linear structures at small scales and perturbations at very large scales. 
This coupling can be represented correctly in cosmological N-Body simulations once halos are resolved at small scales.  
Therefore, we can expect simulations to match our expectations once the scale of non-linearity exceeds the mass resolution scale by more than an order of magnitude. 
At earlier times, there are likely to be errors in the resulting distribution of particles, and these may impact different indicators like power spectrum, mass function, etc., differently.

These expressions provide a firm footing for the ideas of the renormalizability of gravitational clustering in an expanding universe \citep{carrasco2012,couchman1998,crocce2006,hui1996,nishimichi2016,rampf2023}.

\begin{figure*}
    \begin{center}
        \includegraphics[width=0.94\textwidth]{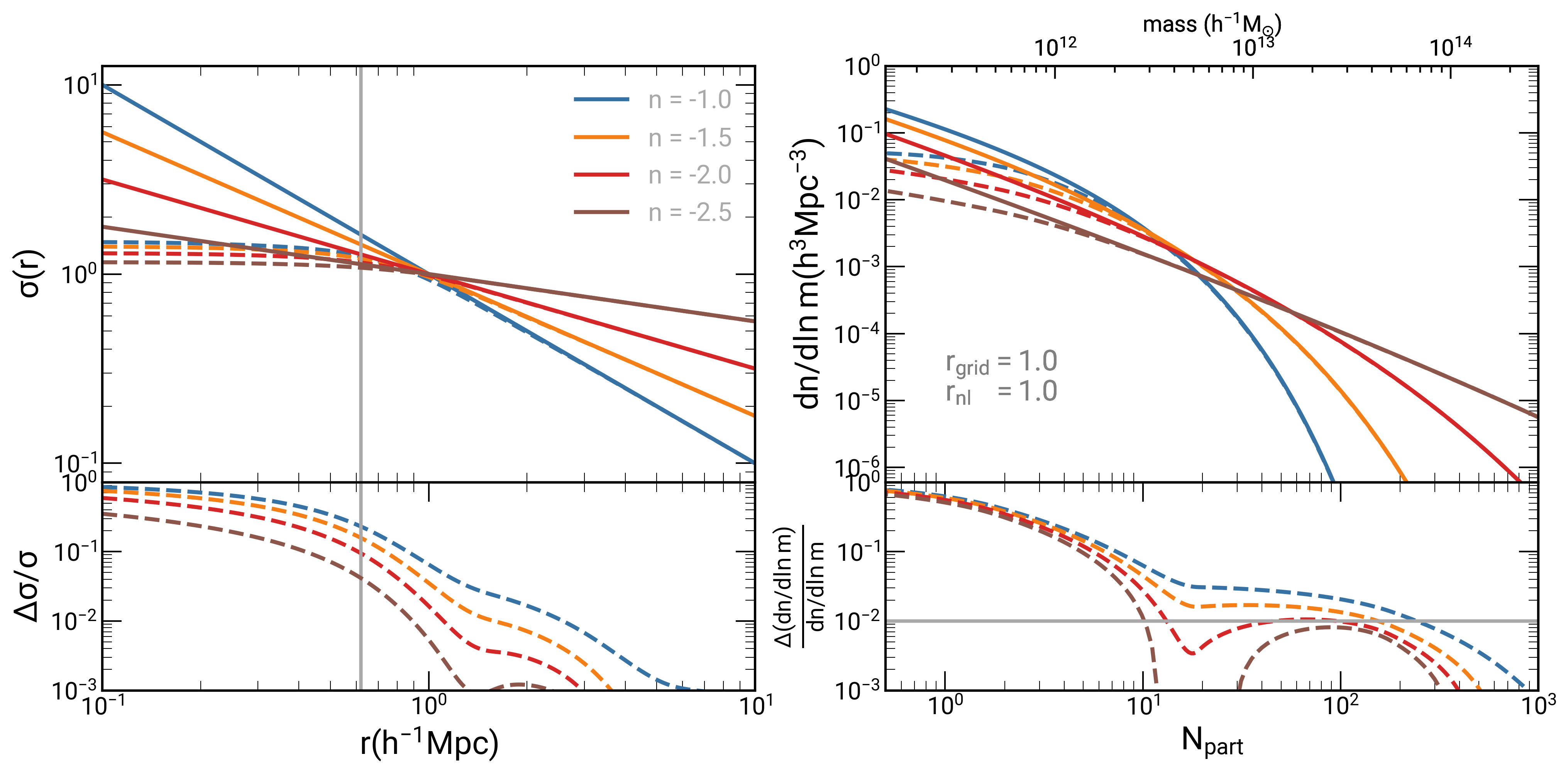}\\
        \caption{Top left panel show mass variance for the power-law power spectrum of indices n. Solid lines are computed using full power spectrum, while dashed lines include constrained power spectrum up to Nyquist frequency. The bottom left panel show fractional errors in mass variance. The vertical grey line shows the size of a single particle. The top right panel slows the halo mass function at a time when the scale of non-linearity is equal to the unit grid length. Solid lines involve a full power spectrum, while dashed lines include a constrained power spectrum. The bottom right panel show fractional errors in mass function. Horizontal grey lines represent the $1$\% mark.}
        \label{fig:th1}
    \end{center}
\end{figure*}

\begin{figure*}
    \begin{center}
        \includegraphics[width=0.94\textwidth]{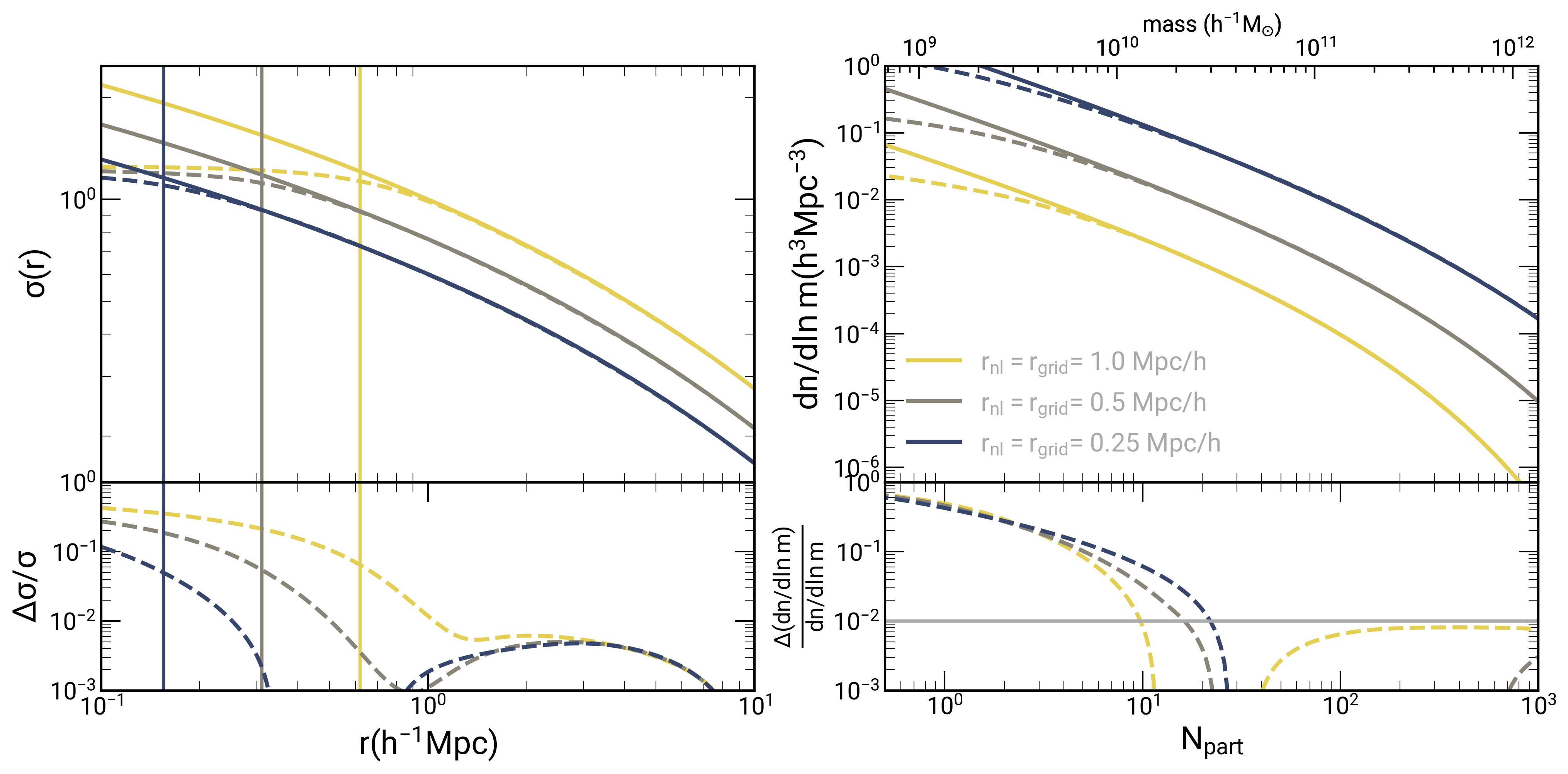}\\
        \caption{The top panels show mass variance and mass function in the $\Lambda$CDM models with three values of $r_{\text{nl}} = r_{\text{grid}}=1.0$, $0.5$, $0.25$ Mpc/h. Solid lines are computed using full power spectrum, while dashed lines include constrained power spectrum up to Nyquist frequency. The bottom panels show fractional errors in mass variance and mass function. The vertical lines in the left panels mark the size of a single particle. Horizontal grey lines represent the $1$\% errors.}
        \label{fig:th_lcdm}
    \end{center}
\end{figure*}

\section{Sources of errors at small scales}
\label{sml_err}

Cosmological N-Body simulations have many potential sources of errors in the results at small scales. We have seen an estimate of the mode coupling between small scale virialised structures and large scale density fluctuations. Any errors or unrealistic representation of density fluctuations and halos at small scales will lead to such errors at large scales via mode coupling. The foremost source of errors at small scales is due to discreteness: particles in a simulation have a finite mass and it is not possible to use these to describe the density field at mass scales that are comparable to or smaller than this mass. Indeed, the density field can only be defined at mass scales much larger than the mass of an individual N-Body particle. Further, simulations cannot resolve collapsed and virialised halos of masses lower than (at least) tens of N-Body particles.  

To be realistic, we also need to worry about errors in inter-particle interaction force, but we drop this aspect as it depends on the specific code being used. 

A corollary of the inability to represent the density field at small mass scales is that no initial perturbation is imprinted in the density distribution at these scales. This missing power translates into modified {\it rms} fluctuations at scales comparable to the grid scale in simulations. We can estimate the effect of this missing power on the mass function of halos using the approach where the {\it rms} fluctuations computed from the power spectrum realised in the initial conditions is used \citep{2006MNRAS.370..993B} and compared with the expected mass function without this limitation. Figure~\ref{fig:th1} shows the expected outcome of such a calculation. Figure~\ref{fig:th1} shows the shift in {\it rms} fluctuations for power law models in the left panel and the corresponding shift in the mass function in the right panel. Here, we have chosen to plot these for the epoch where the scale of non-linearity is one grid length. The lower panels show fractional error as a result of the missing power. We notice that the {\it rms} fluctuations have an error that is larger for models with a larger $(n+3)$, with $n$ being the index of the power spectrum. For some models, the errors in the  {\it rms} exceed $1\%$ out to many grid cells. The corresponding errors in the mass function are also significant, and it is clear that for some models, the error exceeds $1\%$ for halos up to a few hundred particles. 
Figure~\ref{fig:th_lcdm} displays the case $\Lambda$CDM cosmology. The plots are presented for the epoch at which the non-linearity scale equals one grid length. As the effective spectral index is such that $(n_{\text{eff}}+3)$ is small at very small scales in this model, the errors drop to 1\%  very quickly and hence the missing power is not a serious problem for $\Lambda$CDM models.

\begin{figure*}
\begin{center}
\begin{subfigure}[b]{0.32\textwidth}
    \includegraphics[width=0.99\textwidth]{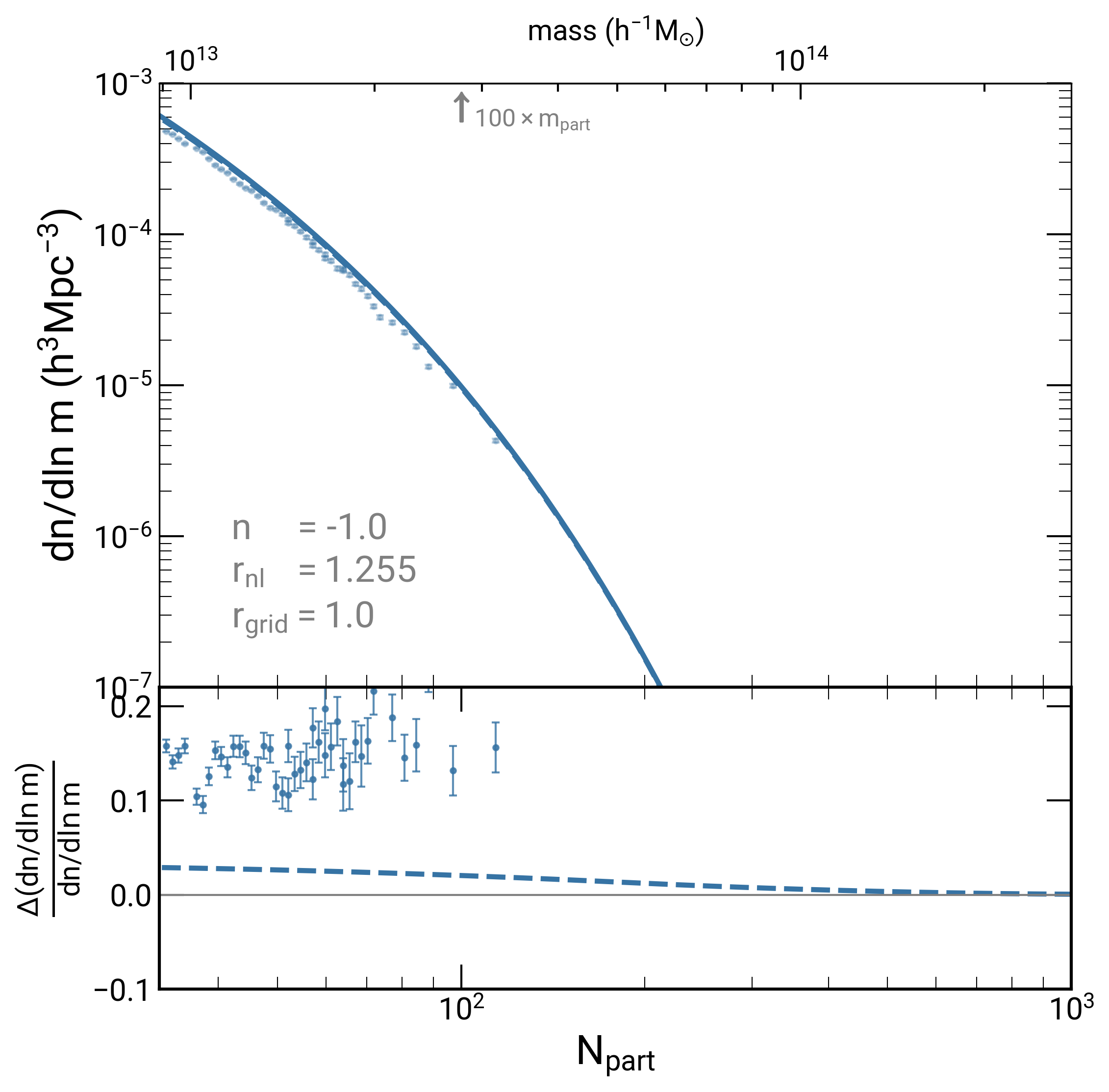}
\end{subfigure} 
\begin{subfigure}[b]{0.32\textwidth}
    \includegraphics[width=0.99\textwidth]{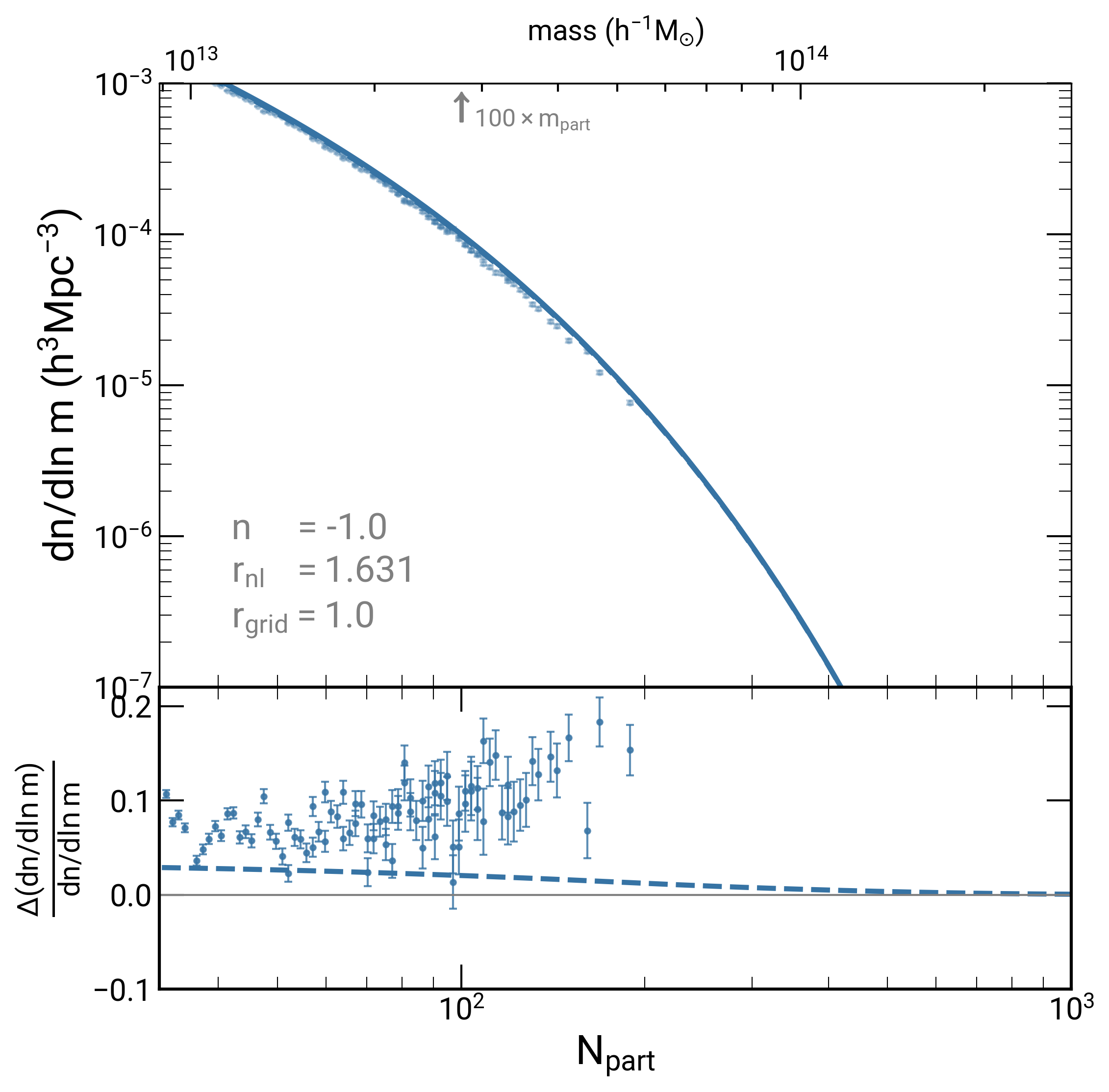}
\end{subfigure} 
\begin{subfigure}[b]{0.32\textwidth}
    \includegraphics[width=0.99\textwidth]{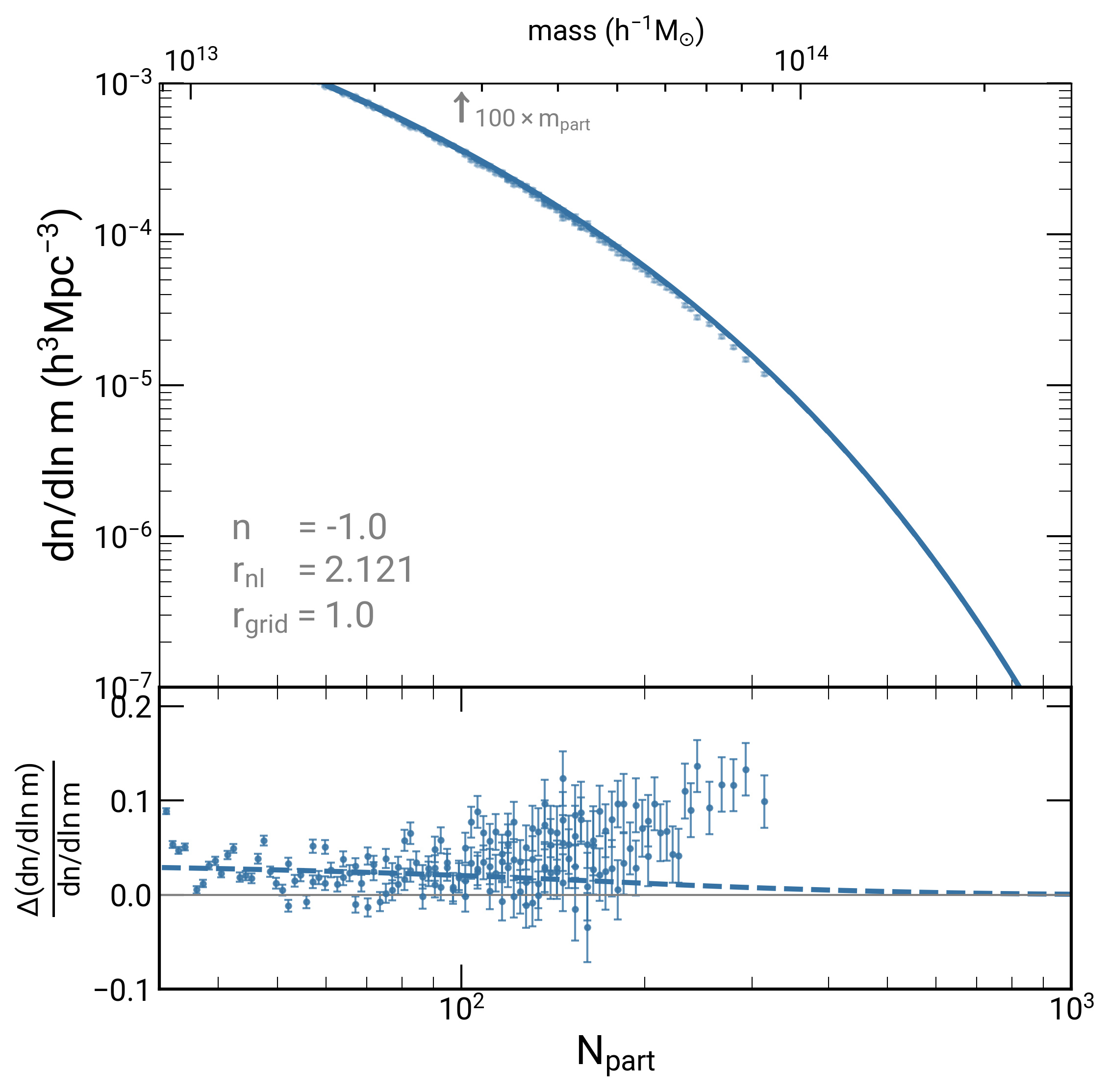}
\end{subfigure} \\
\begin{subfigure}[b]{0.32\textwidth}
    \includegraphics[width=0.99\textwidth]{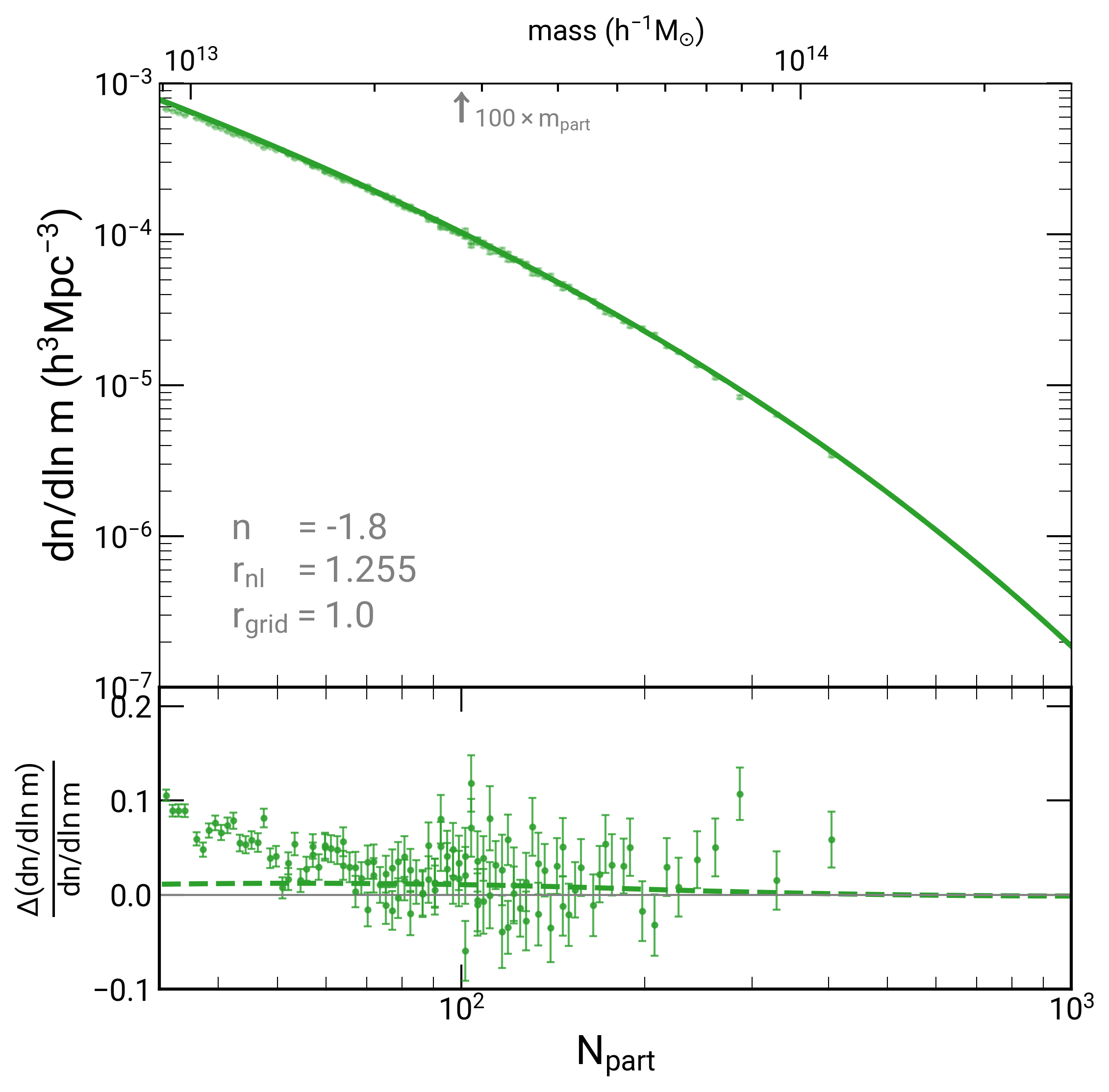}
\end{subfigure} 
\begin{subfigure}[b]{0.32\textwidth}
    \includegraphics[width=0.99\textwidth]{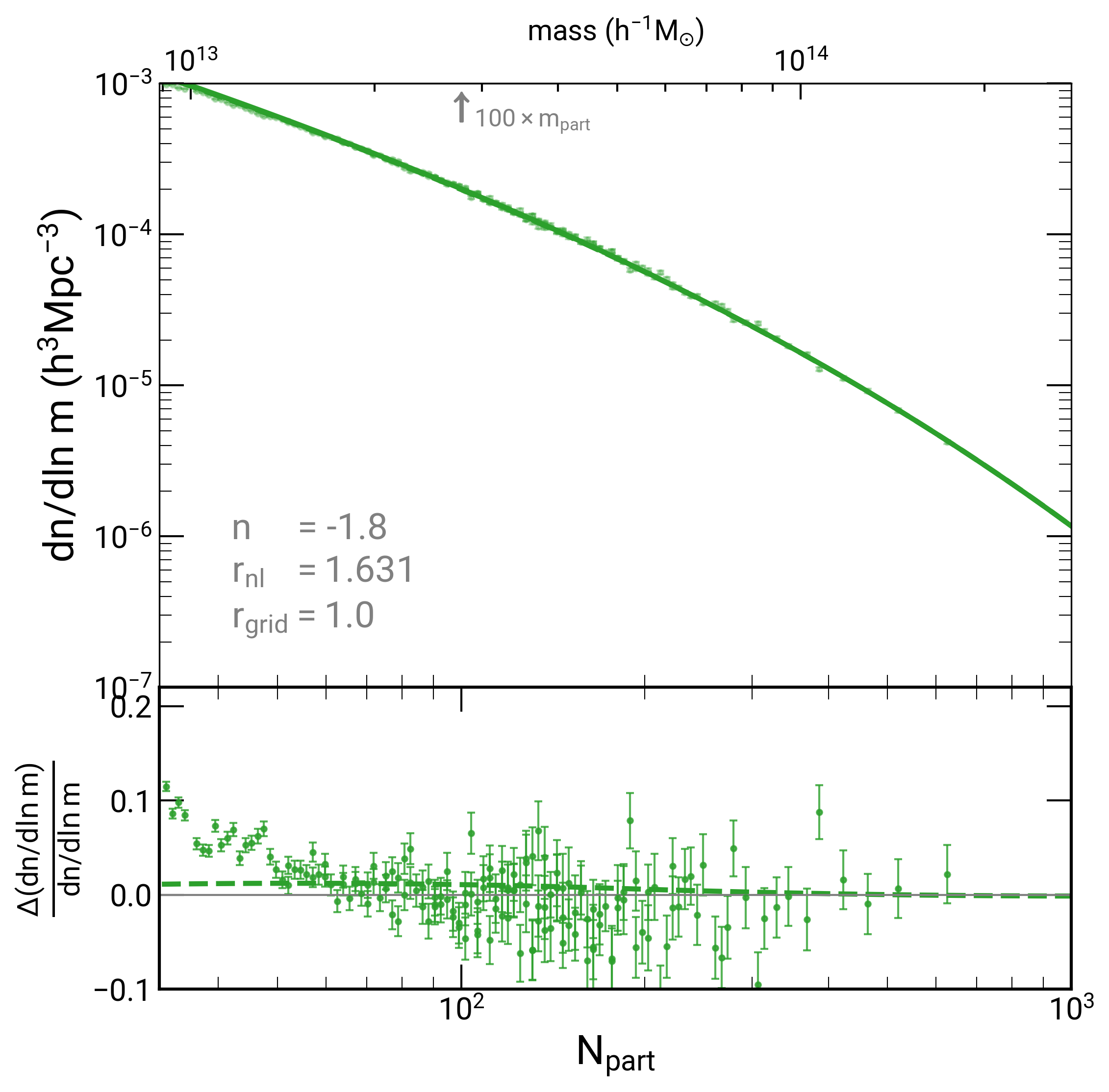}
\end{subfigure} 
\begin{subfigure}[b]{0.32\textwidth}
    \includegraphics[width=0.99\textwidth]{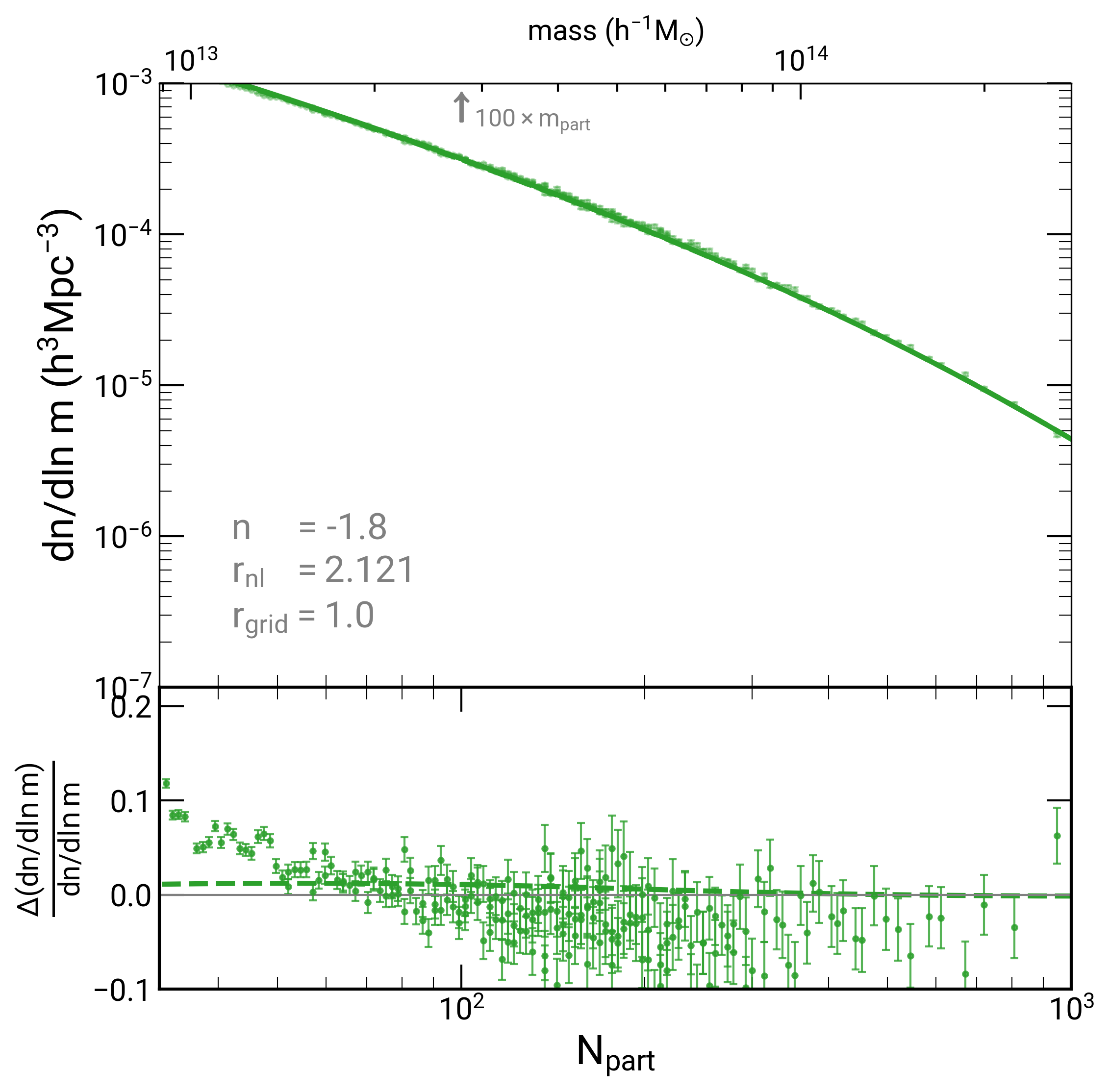}
\end{subfigure} \\
\begin{subfigure}[b]{0.32\textwidth}
    \includegraphics[width=0.99\textwidth]{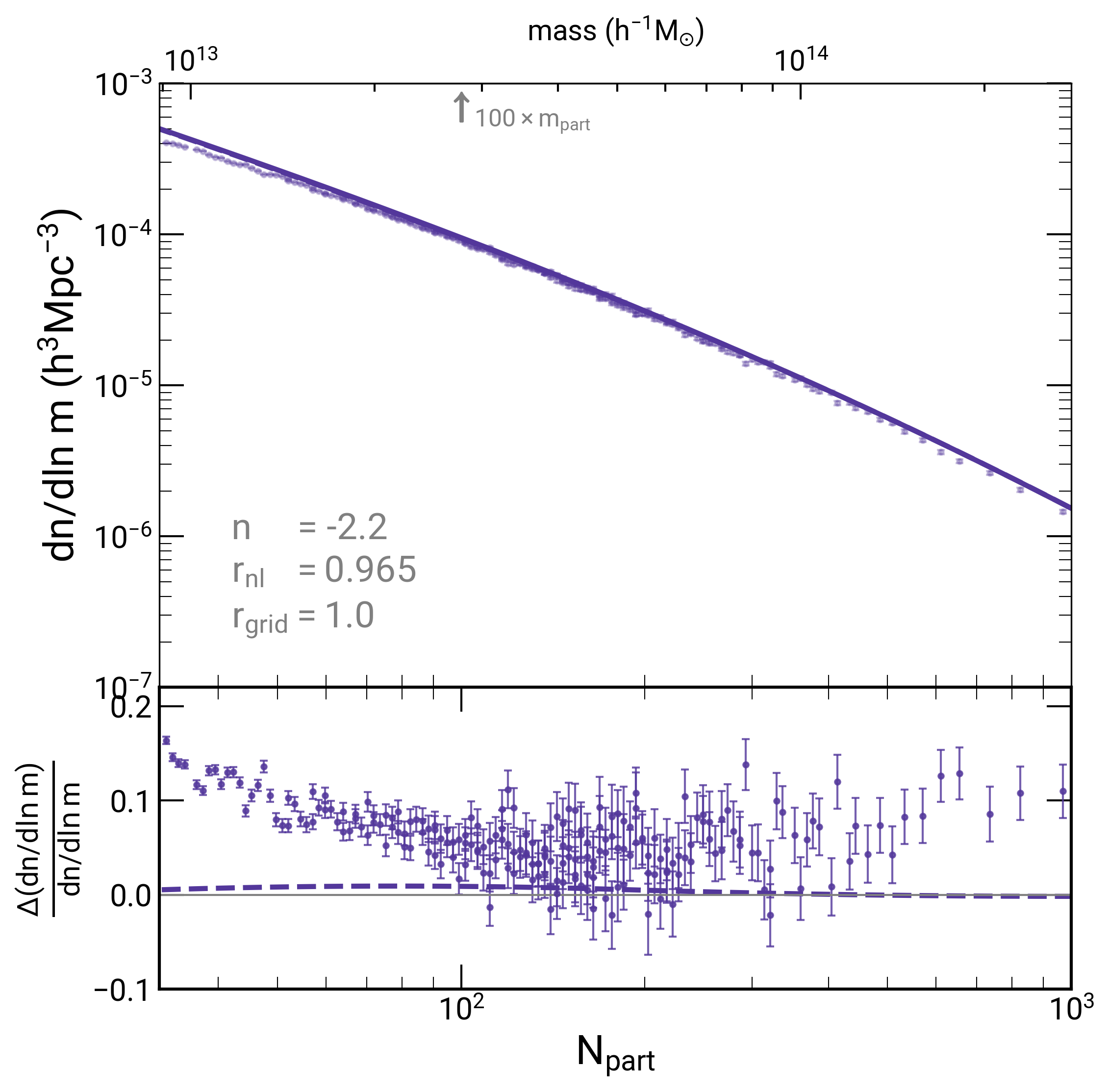}
\end{subfigure} 
\begin{subfigure}[b]{0.32\textwidth}
    \includegraphics[width=0.99\textwidth]{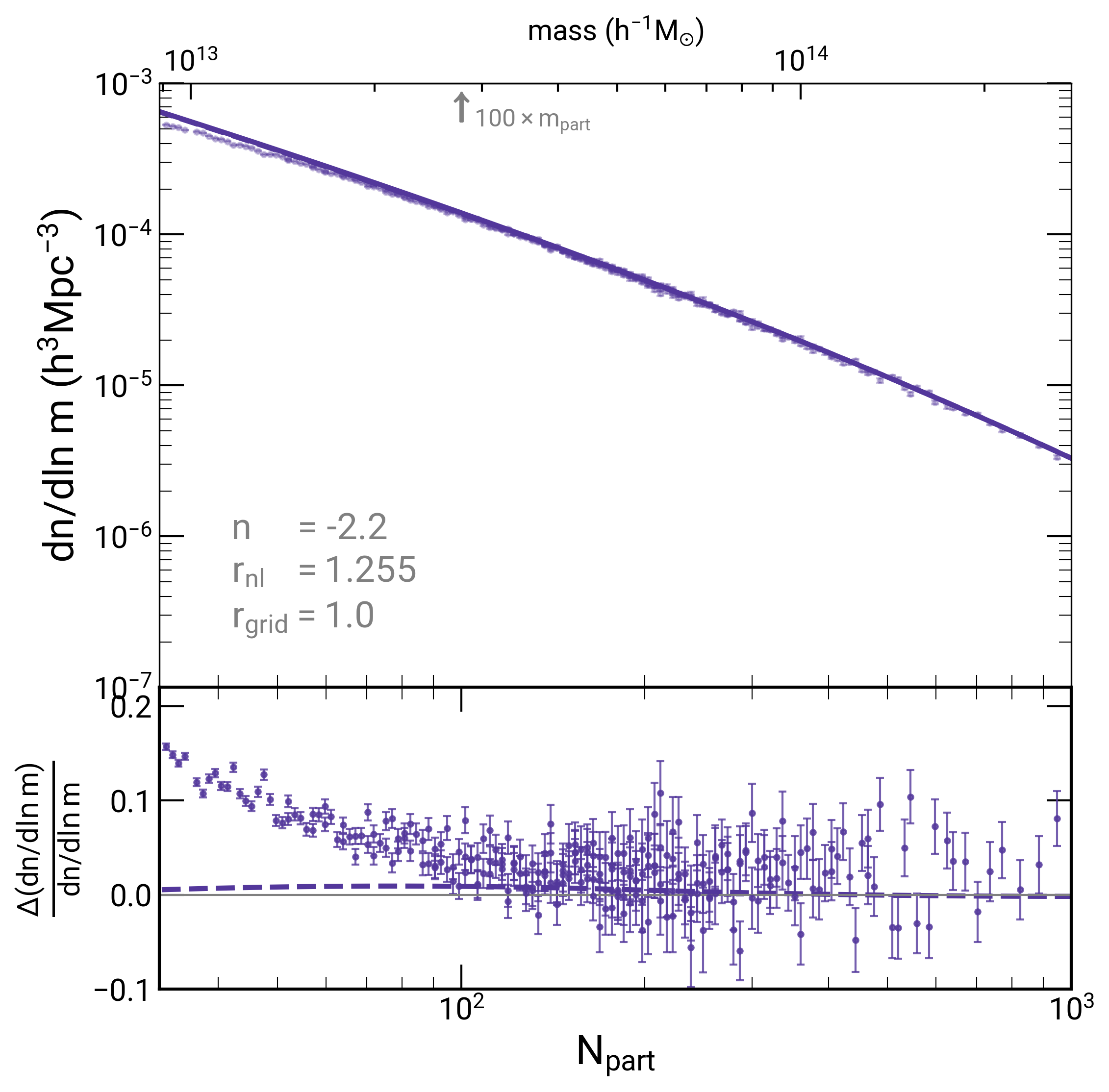}
\end{subfigure} 
\begin{subfigure}[b]{0.32\textwidth}
    \includegraphics[width=0.99\textwidth]{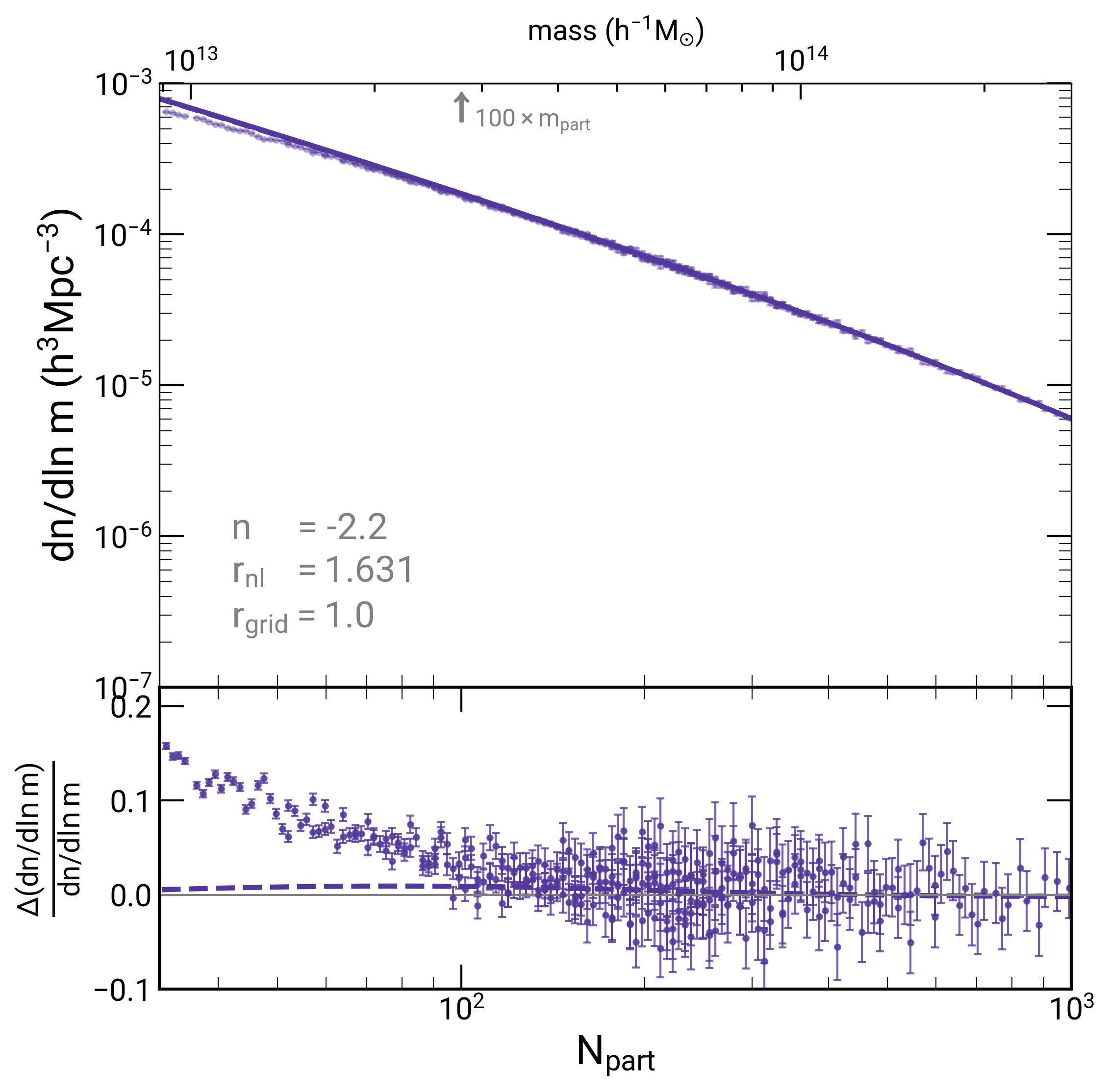}
\end{subfigure}
\caption{Each panel show the halo mass function and fractional errors in mass function as a function of a number of particles. Data points represent simulation, whereas solid lines show theoretical mass function. The top, middle and bottom rows represent the power-law power spectrum index $n$ values -1.0, -1.8 and -2.2, respectively. Columns represent different scales on non-linearity ($r_{nl}$). }
\label{fig:sim}
\end{center}
\end{figure*}

Figure~\ref{fig:sim} shows the mass function for three models at three epochs each.  Each row is for a given index: the top row is for $n=-1$, the middle row is for $n=-1.8$, and the bottom row is for $n=-2.2$. Each row has plots for three epochs corresponding to an increasing scale of non-linearity as we go from left to right. Each panel shows the expected mass function shown in eq. \ref{eqn:mf} and the mass function from N-Body simulations. The lower part of each panel shows the fractional error in the mass function. We note that the error is indeed highest at early epochs for $n=-1$, which corresponds to the largest $(n+3)$ in this set. This is in agreement with the impact of the missing power at small scales estimated above. We see that the errors come down over a substantial range in mass at later epochs. Some of the remaining disagreement results from a poor fitting function for the theoretical estimate, discussed in \cite{Gavas_2023}. Figures 3, 4, and 5 of \cite{Gavas_2023} illustrate the poor fit of the halo mass function to scale-free simulations. Using the Sheth-Tormen functional form \citep{Sheth_2001} to fit their data, they provide the Sheth-Tormen parameters ($p$, $q$) as a function of power-law power spectrum index $n$, which we utilize for figure \ref{fig:sim}.
\begin{subequations}
\label{eqn:mf}
\begin{align}
\frac{dn}{d \ln m} &= \frac{\bar{\rho}}{m} \frac{d \ln \sigma^{-1}}{d \ln m} f(\nu), \\
f(\nu) &= A(p) \nu \sqrt{\frac{2q}{\pi}} [1+(q\nu^2)^{-p}]  \exp(-q\nu^2 /2),\\
  p(n) &= -0.045 \ n+0.231 ,\\ 
  q(n) &= 0.095 \ n+0.922,
\end{align}
\end{subequations}
where, $\nu$ is peak height $\nu \equiv \delta_c /(D_+(z)\sigma(m))$. Here $\delta_c$ is the linearly extrapolated critical over density required for a spherically symmetric perturbation. $D_+$ is the growth factor. $\sigma(m)$ is the variance in the initial density fluctuation field, linearly extrapolated to the present epoch and smoothed with a top hat filter $W(k,m)$ of scale R= $(3m/4\pi\bar{\rho})^{1/3}$.

\section{Discussion and Summary}

We have seen in the last two sections that the mode coupling between small scales and large scales has a component arising from interactions of halos. This mode coupling arises from the departure of shapes of halos from spherical shapes and tidal interactions. These components clearly require the halos to be resolved in the simulations with a large number of particles. Thus, we expect that this is a source of errors in cosmology N-Body simulations and that it arises from discreteness. It is reasonable to expect that the error will diminish at late times once the scale of non-linearity is sufficiently larger than the average inter-particle separation. This is motivated by observations that clustering at small scales is dominated by the power spectrum around the scale of non-linearity \cite{little1991,bagla1997b}.

The second effect on the limitation of N-Body simulations at small scales and at early times arises from missing power. It is not possible to represent density perturbations below the Nyquist frequency, and hence, there is a sharp departure from the scenario being simulated. We show that this results in an underestimation of the {\it rms} fluctuations at small scales in the simulation. Obviously, this affects models with steep power spectra much more than other models. We quantify this by computing the {\it rms} amplitude of fluctuations and the corresponding mass function of collapsed halos. We find that the errors percolate to an under-estimated {\it rms} amplitude of fluctuations out to several grid lengths and mass function for halos with more than $10^2$ particle halos at early times. 

We test this deficit in the mass function using N-Body simulations and find that this is seen clearly at early times for the $n=-1$ power spectrum. We also note that the error diminishes at later times. The error is smaller for other models for halos with $50$ or more particles, even at early times. 

The summary of our findings is that there are significant errors in cosmological N-Body simulations at small scales at early times. In general, the results of simulations are not reliable till the scale of non-linearity is larger than one grid length. For steeper power spectra with an index where $(n+3) > 1$, it may be better to only use data once the scale of non-linearity is much larger than a grid length. Note that we have not considered any errors that may arise due to errors in force or numerical implementation, e.g., \cite{1985A&A...144..413B,bagla1997}. These limitations on cosmological N-Body simulations are complementary to the limitations arising from a finite box size \cite{2006MNRAS.370..993B}.  

A comparison with some of the other estimates of expected transients is also in order.  \cite{Scoccimarro_1998} estimated transients arising from the deviations of initial conditions generated using the Zel'dovich approximation as compared to the expected evolution.  Note that the Zel'dovich approximation gives correct evolution at linear order and hence any deviations arise from its use to set up initial conditions directly in the mildly non-linear regime.  This estimation was done using higher order Lagrangian perturbation theory.  We would like to point out that we have assumed that the generation of initial conditions as well as dynamical evolution is {\sl ideal} within the limitations of mass resolution and the range of wave modes available in the simulation.
Interestingly, analytical calculations using a gravitational dynamics approach also indicate a spectrum dependence in the influence of small scales on larger scales \cite{1996ApJ...472....1B} where a similar factor of $(n+3)$ makes an appearance\footnote{See end of \S5 in \cite{1996ApJ...472....1B}.}

The primary concern raised in this work is that N-Body simulations are plagued by transient features introduced by two factors: discreteness induced errors in mode coupling and, missing power at small scales.  We do not dwell on any other {\sl non-ideal} aspects of evolution of perturbations in cosmological N-Body simulations.  As we have shown, the errors arising out of these factors are significant at small scales and at early epochs where the scale of non-linearity is smaller than a few grid lengths.  Once the scale of non-linearity is such that halos with tens of particles have collapsed and virialised, these errors become insignificant.  Thus we are pointing out that the reliability of cosmological N-Body simulations is limited at smaller scales.  When combined with the limitations arising from box size (See, e.g., \cite{bagla2005}), we see that the dynamic range of cosmological simulations in length scales is about an order of magnitude smaller than the box size (in terms of number of cells).  As these limitations cannot be overcome within the context of simulations, other approaches that help enhance the dynamic range of simulations must be studied. These results imply that the use of very large simulations to generate mock catalogues needs to be evaluated carefully and suitable corrections devised in order to ensure that these are reliable, e.g., see \cite{Ramakrishnan_2021}. Of course, as all such methods are calibrated with simulations, sufficient care and caution needs to be exercised.  Approaches that suffice for some quantities of interest may fail for other observables and hence the validation of these methods has to be carried out with some care before deploying these for real applications.

One may wonder whether such errors arising at small scales will contribute to larger scales at late time. 
However, we know from many numerical experiments that once larger scales become non-linear these can erase differences at smaller scales \cite{little1991,bagla1997b}.

We may summarise the lessons from this work as follows: N-Body simulations can be trusted for generic CDM/LCDM type models once the scale of non-linearity exceeds the mass resolution by at least an order of magnitude.  
For such situations the {\it correctness} of cosmological N-Body simulations and renormalizability of gravitational clustering can be justified. 

\section*{Acknowledgements}

JSB would like to thank Rajaram Nityananda, T. Padmanabhan, K. Subramanian and Nishikanta Khandai for several discussions on this subject.
JSB also thanks NCRA-TIFR for hospitality, as this manuscript was completed during the sabbatical from IISER Mohali. 
SG extends sincere thanks to Pierluigi Monaco and Stephane Colombi for useful discussions and suggestions.
This research has made use of NASA's Astrophysics Data System.

\vspace{-1em}

% \bibliographystyle{unsrt}

% \bibliography{main}

\appendix

\section{Simulations}
\label{a1}
In this section, we describe the simulations used in \S~\ref{sml_err} and Figure~\ref{fig:sim}. We also demonstrate the convergence of the mass function computed in these simulations.

\begin{table*}[h]
	\begin{center}
	 \begin{tabular}{|c c c c c c c|} 
	 \hline 
	 n & $z_{\text{start}}$ & $N_{\text{box}}$ & $\sqrt[3]{N_{\text{part}}}$  & $r_{nl}^{\text{sim}}$ & $r_{nl}^{\text{mf}}$ & $r_{nl}^{\text{max}}$ \\ [0.5ex] 
	 \hline
	 -1.0 & 160 & 512 & 512  & 1.25-49.4& 1.265, 1.631, 2.121 & 27.4 \\[0.5ex]
	 -1.0 & 160 & 768 & 768  & 1.25-49.4& 1.265, 1.631, 2.121 & 27.4 \\[0.5ex]
	 -1.0 & 160 & 1024 & 768  & 1.25-49.4& 1.631, 2.121, 2.757 & 27.4 \\[0.5ex]
	 -1.0 & 160 & 1024 & 1024  & 1.25-49.4& 1.265, 1.631, 2.121 & 27.4 \\[0.5ex]
	 -1.8 & 70 & 1024 & 1024  & 1.25-49.4& 1.265, 1.631, 2.121 & 14.5 \\[0.5ex]
	 -2.2 & 46 & 1536 & 1536  & 0.2-10.24& 0.965, 1.265, 1.631 & 4.4 \\[0.5ex]
	 \hline

	 \end{tabular}
	 \caption{\label{tab:sim}{\bf Simulation Setup:} Column 1: Power law power spectrum index for the model, Column 2: Initial redshift used to start the simulation, Column 3: Side length of the cubical simulation box, Column 4: Cube root of the total number of particles put in the simulation, Column 5: Range of the scale of non-linearity($r_{nl}$) covered in the simulation, Column 6: $r_{nl}$s used to compute the mass function, Column 7: Maximum limit on $r_{nl}$ considering the finite box size effect.}
	\end{center}
\end{table*}
    
We use six dark-matter-only simulations initialised with three power-law power spectrum conditions. The power-law power spectrum indices $n$ are listed in Column~1 of Table~\ref{tab:sim}. Simulations are run using the \textsc{GADGET-4} code (\cite{Springel_2021}), configured in TreePM mode, with initial conditions generated via second-order Lagrangian perturbation theory. The softening length is set to $\epsilon_0 = 0.05$ grid units for all runs. Adaptive time steps are utilised with a maximum step size of 0.005. These parameter choices secure that force calculation errors remain well below 1\%. Halos are identified using the \textsc{FoF-SUBFIND} algorithm. The cosmological model adopted is Einstein-de Sitter (EdS), where the $r_\text{nl}$ and scale factor relation is given by:
\begin{equation}
r_\text{nl}  \propto a^{\frac{2}{3+n}}.
\end{equation}
All power spectra are normalised such that $\sigma_\text{lin}(a = 1, r_\text{nl} = 8) = 1$. We account for finite box size effects (\cite{bagla2006}) while selecting the appropriate box size and the maximum value of $r_\text{nl}$ where results can be considered reliable and is shown in column 7 of Table~\ref{tab:sim}. Additional details about the simulations are provided in Table~\ref{tab:sim}.

\begin{figure*}[h]
\begin{center}
\begin{subfigure}[b]{0.32\textwidth}
    \includegraphics[width=0.99\textwidth]{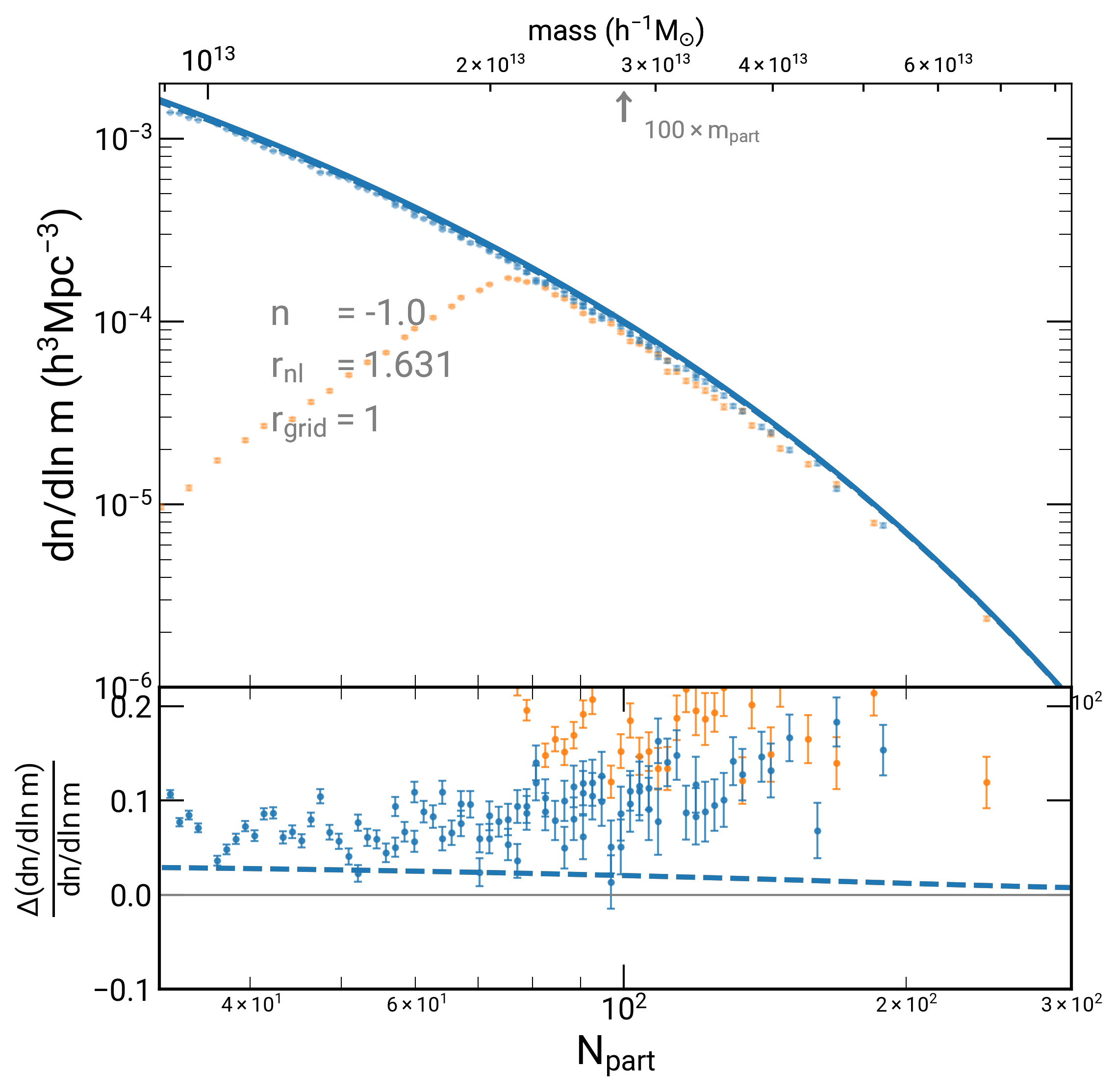}
\end{subfigure} 
\begin{subfigure}[b]{0.32\textwidth}
    \includegraphics[width=0.99\textwidth]{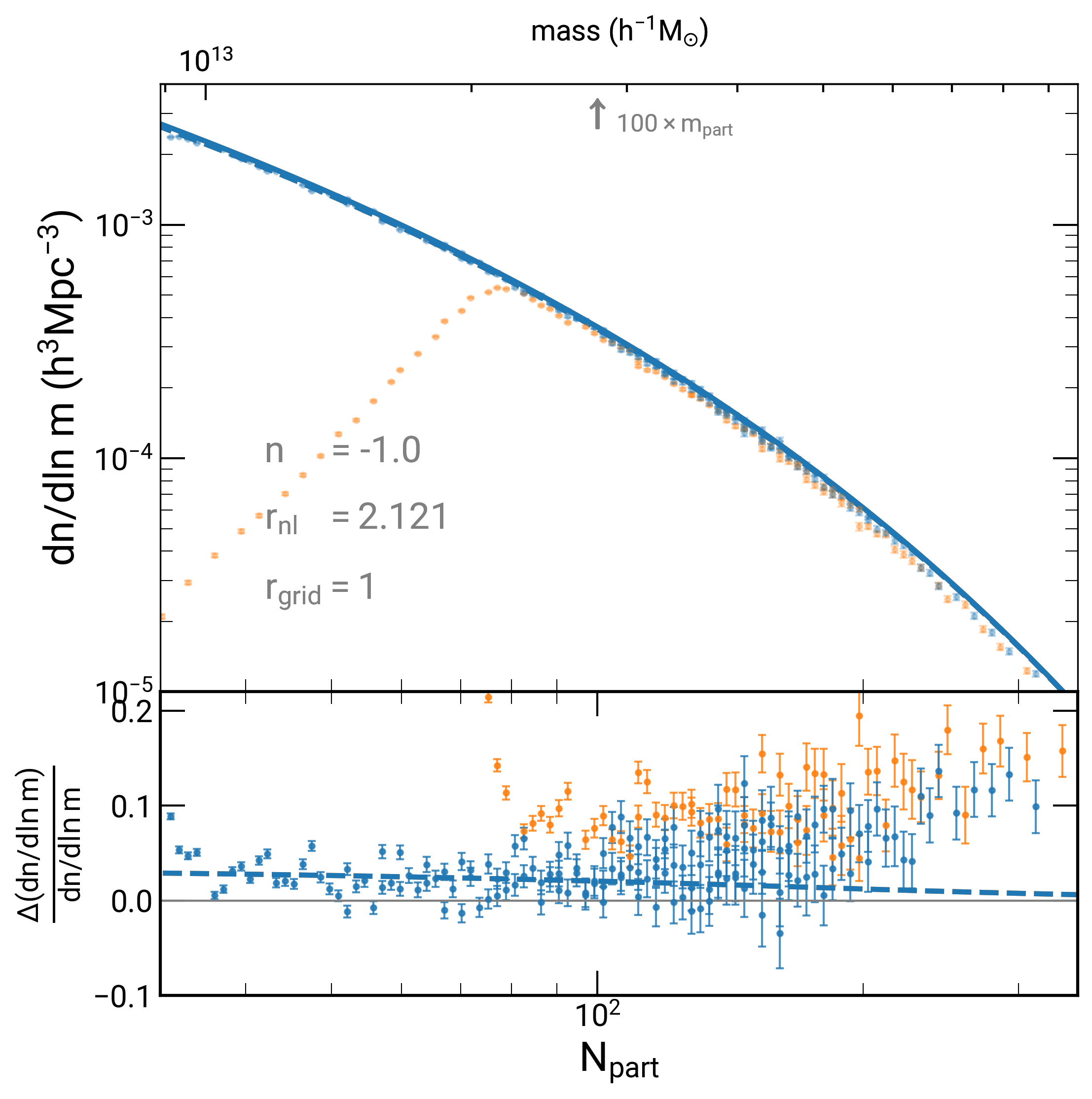}
\end{subfigure} 
\begin{subfigure}[b]{0.32\textwidth}
    \includegraphics[width=0.99\textwidth]{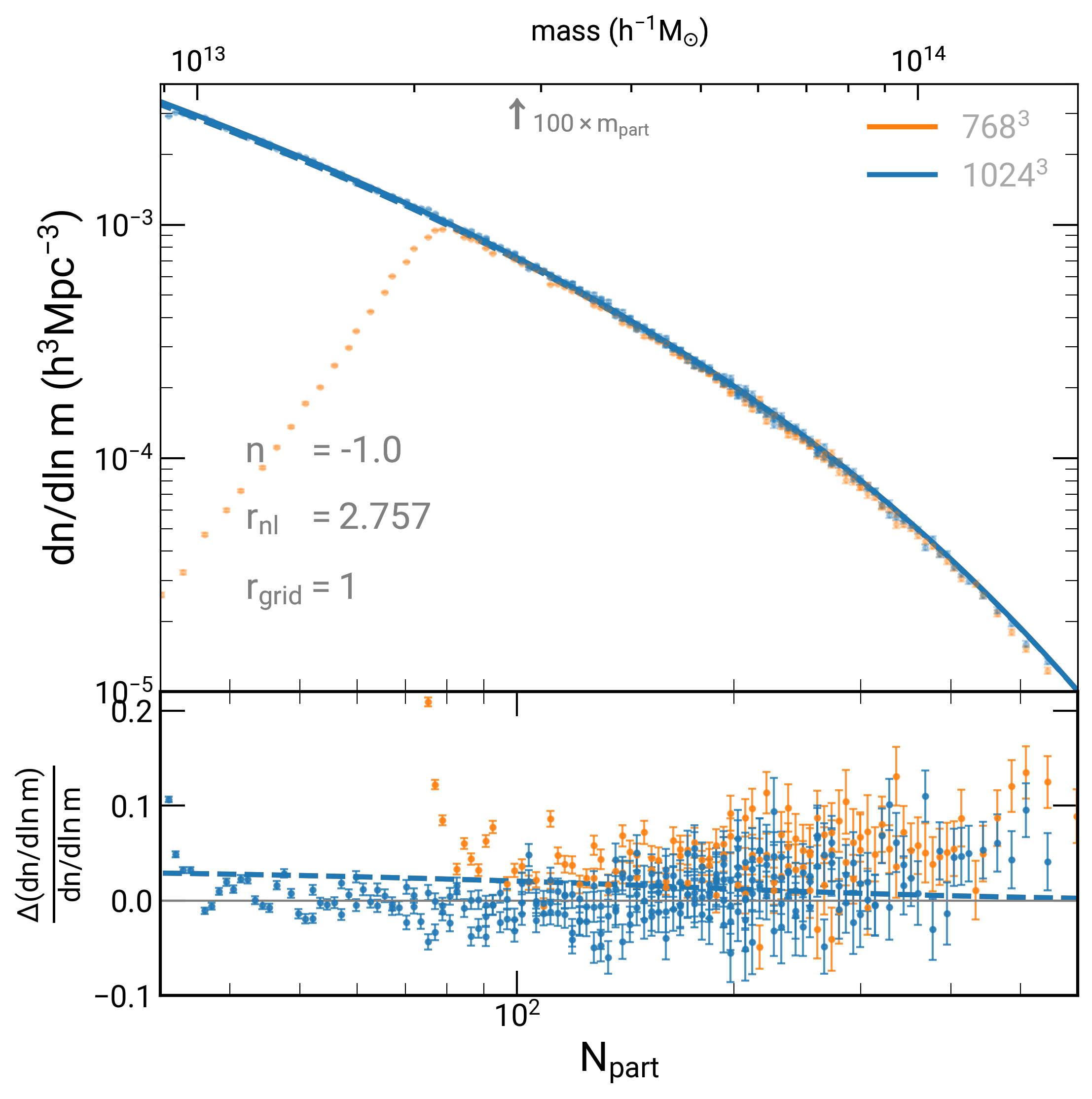}
\end{subfigure} 
\caption{Each panel shows the halo mass function and fractional errors in mass function in the simulation for the $n=-1.0$ model with two values of $N_{\text{part}}= 768^3$ (orange), and $1024^3$ (blue) of box size $1024^3$ shown in Table \ref{tab:sim}. Data points represent simulation, whereas solid lines show theoretical mass function. Three panels represent different scales on non-linearity ($r_{nl}$).}
\label{fig:sim_conv1}
\end{center}
\end{figure*}

\begin{figure*}[h]
\begin{center}
\begin{subfigure}[b]{0.32\textwidth}
    \includegraphics[width=0.99\textwidth]{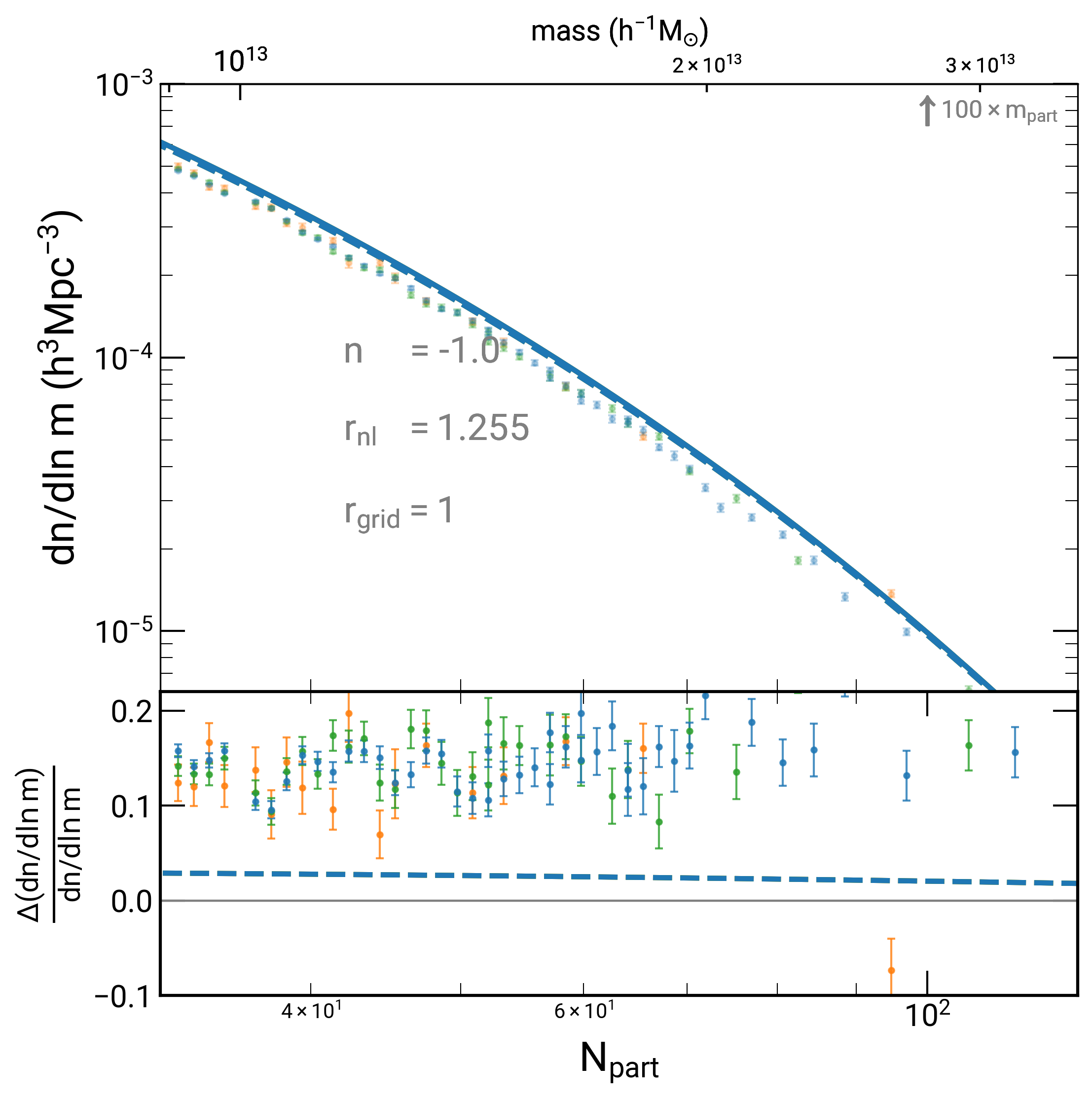}
\end{subfigure} 
\begin{subfigure}[b]{0.32\textwidth}
    \includegraphics[width=0.99\textwidth]{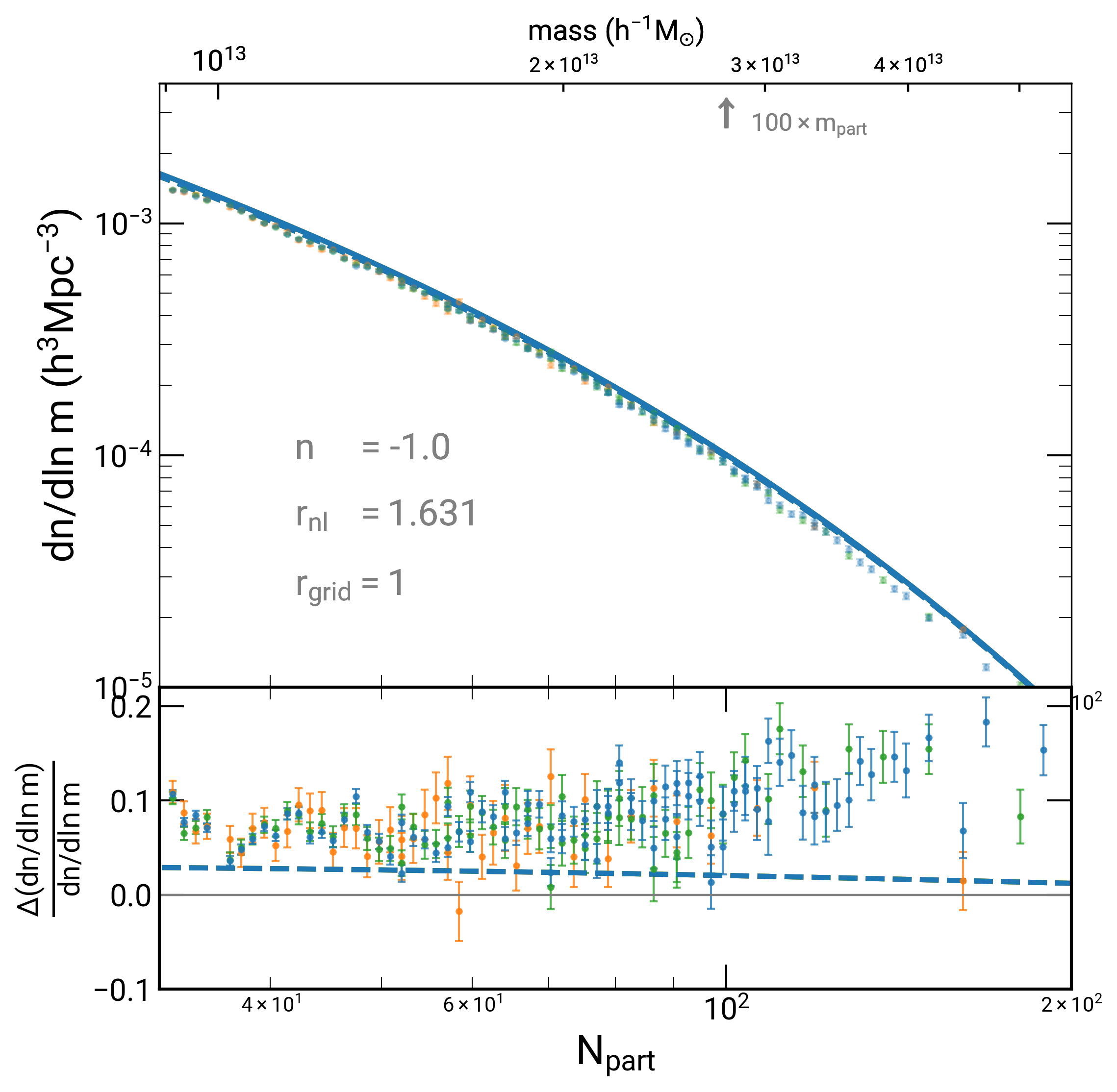}
\end{subfigure} 
\begin{subfigure}[b]{0.32\textwidth}
    \includegraphics[width=0.99\textwidth]{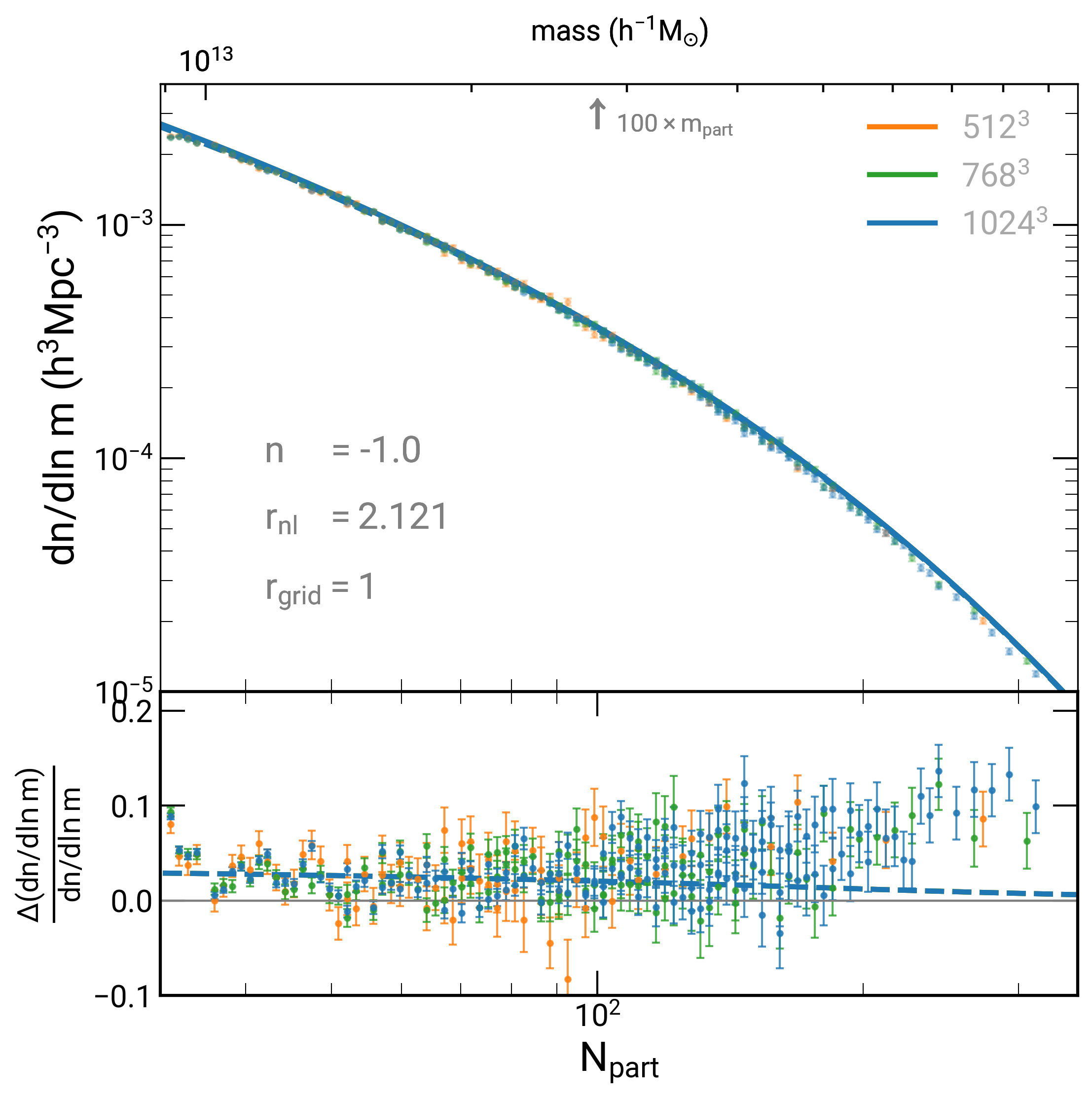}
\end{subfigure} 
\caption{Each panel shows the halo mass function and fractional errors in mass function in the simulation for three variations of $N_{\text{part}}=N_{\text{box}}^3=$ $512^3$ (orange), $768^3$ (green), $1024^3$ (blue) for the $n=-1.0$ model shown in Table \ref{tab:sim}. Data points represent simulation, whereas solid lines show theoretical mass function. Three panels represent different scales on non-linearity ($r_{nl}$).}
\label{fig:sim_conv2}
\end{center}
\end{figure*}

Figure~\ref{fig:sim_conv1} shows a conversion of the mass function computed by the simulation with respect to finite mass resolution. The lower-resolution scatters show higher fractional errors in mass function at a given $r_{nl}$. However, these errors settle similarly with the $r_{nl}$. Similarly, Figure~\ref{fig:sim_conv2} displays the impact of finite box size on the mass function. The box size affects the range of masses that can be probed, with larger box sizes allowing for better coverage of the higher-mass end.

\end{document}